\documentclass[epj]{svjour}
\usepackage{graphicx}
\usepackage{color}

\def\mb#1{\setbox0=\hbox{$#1$}\kern-.025em\copy0\kern-\wd0
\kern-0.05em\copy0\kern-\wd0\kern-.025em\raise.0233em\box0}

\begin{document}
   \title{Self-gravitating Brownian particles in two dimensions:\\
    the case of $N=2$ particles}

 \author{P.H. Chavanis and R. Mannella}

\institute{$^1$ Laboratoire de Physique Th\'eorique (IRSAMC), CNRS and UPS, Universit\'e de Toulouse, F-31062 Toulouse, France\\$^2$  Dipartimento di Fisica `E.
Fermi' (CNISM Unit\`{a} di Pisa), CNR-INFM, Universit\`{a} di Pisa, Largo Pontecorvo 3, \\ 56127 Pisa, Italy\\
\email{chavanis@irsamc.ups-tlse.fr; mannella@df.unipi.it}
}

\titlerunning{Self-gravitating Brownian particles in two dimensions}

   \date{To be included later }

   \abstract{We study the motion of $N=2$ overdamped Brownian
   particles in gravitational interaction in a space of dimension
   $d=2$. This is equivalent to the simplified motion of two
   biological entities interacting via chemotaxis when time delay and
   degradation of the chemical are ignored. This problem also bears
   some similarities with the stochastic motion of two point vortices
   in viscous hydrodynamics [Agullo \& Verga, Phys. Rev. E, {\bf 63},
   056304 (2001)].  We analytically obtain the probability density of
   finding the particles at a distance $r$ from each other at time
   $t$. We also determine the probability that the particles have
   coalesced and formed a Dirac peak at time $t$ (i.e. the probability
   that the reduced particle has reached $r=0$ at time $t$). Finally,
   we investigate the variance of the distribution $\langle
   r^2\rangle$ and discuss the proper form of the virial theorem for
   this system. The reduced particle has a normal diffusion behavior
   for small times with a gravity-modified diffusion coefficient
   $\langle r^2\rangle=r_{0}^2+(4k_B/\xi\mu)(T-T_{*})t$, where
   $k_BT_{*}=Gm_1m_2/2$ is a critical temperature, and an anomalous
   diffusion for large times $\langle r^2\rangle\propto
   t^{1-T_*/T}$. As a by-product, our solution also describes the
   growth of the Dirac peak (condensate) that forms at large time in
   the post collapse regime of the Smoluchowski-Poisson system (or
   Keller-Segel model) for $T<T_c=GMm/(4k_B)$. We find that the
   saturation of the mass of the condensate to the total mass is
   algebraic in an infinite domain and exponential in a bounded
   domain. Finally, we provide the general form of the virial theorem
   for Brownian particles with power law interactions.
 }

   \maketitle
%

\section{Introduction}
\label{sec_model}

Systems with long-range interactions have recently been the object of
considerable interest \cite{cdr}. One usually considers isolated
systems in which the particles evolve according to deterministic
Hamiltonian equations. These systems are described by the
microcanonical ensemble. Examples of such systems include
self-gravitating systems, two-dimensional point vortices, the
Hamiltonian mean field (HMF) model etc. However, one may also consider
dissipative systems in which the particles, in contact with a thermal
bath, evolve according to stochastic Langevin equations. These systems
are described by the canonical ensemble. The statistical mechanics of
Hamiltonian and Brownian systems with long-range interactions is
discussed in \cite{hb2} at a general level. In this paper, we consider
the case of Brownian particles in gravitational interaction. It is
known that this system bears deep analogies with simple models of
bacterial populations experiencing chemotaxis in
biology\footnote{These systems are isomorphic up to a change of
notations. In this paper, we shall use the notations of astrophysics
because they are closer to the notations that are familiar in physics
and thermodynamics. However, our results can be transposed easily to
the biological context. We refer to Perthame
\cite{perthame} for a complete bibliography of the  chemotactic problem
from the viewpoint of applied mathematics and to Chavanis \& Sire for additional references in physics \cite{kinbio}.} (see, e.g.,
Ref. \cite{crrs} for a description of this
analogy). In a proper thermodynamic limit $N\rightarrow +\infty$, the
mean field approximation becomes exact and the dynamics of these
systems is described by the Smoluchowski-Poisson system (gravity)
\cite{virial} or by the Keller-Segel model (chemotaxis)
\cite{ks}. These equations display rich phenomena such as collapse and
evaporation.  In particular, in $d=2$ dimensions, the evolution leads
to the formation of Dirac peaks if the temperature is below the
critical value $k_BT_c=GMm/4$ \cite{herrerobio,sc,lushnikov}. These
systems have been studied essentially in the mean field approximation,
i.e. for a large number of particles. The case of a finite number of
particles can be studied numerically by solving the $N$-body
stochastic equations. Numerical results will be presented in a
companion paper
\cite{mc}. In the present paper, we consider the extreme case of only
$N=2$ Brownian particles in gravitational interaction that can be
solved analytically.  We analytically obtain the probability density
of finding the particles at a distance $r$ from each other at time $t$
and determine the probability that the particles have coalesced and
formed a Dirac peak at time $t$. We also investigate the variance of
the distribution $\langle r^2\rangle$ and discuss the proper form of
the virial theorem for this system. In particular, we show that the
virial theorem obtained in
\cite{epjb} is only valid as long as the particles have not formed
Dirac peaks.

The paper is organized as follows. In Sec. \ref{sec_position} we
recall the $N$-body coupled stochastic equations describing the
evolution of self-gravitating Brownian particles and specifically
consider the case $N=2$. We introduce the center of mass and the
reduced particle. We show that the center of mass undergoes a pure
Brownian motion and that the reduced particle undergoes a Brownian
motion in a central potential $U=Gm_1m_2\ln r$. We also recall the
``naive'' virial theorem obtained in \cite{epjb} and discuss, with a
new light, the distinction between the critical temperatures
$k_BT_c=Gm_1m_2/4$ and $k_BT_*=Gm_1m_2/2$. In Sec. \ref{sec_bn}, we
study the motion of a Brownian particle (reduced particle) in an
attractive central potential $U=Gm_1m_2\ln r$ in $d=2$. We show that
the corresponding Fokker-Planck equation is equivalent to a
Schr\"odinger equation (with imaginary time) with a potential
$V=-D(a/r)^2$. This equation can be solved analytically in terms of
Bessel functions. Then, we can obtain various analytical results such
as the probability to find the reduced particle at position $r$ at
time $t$, the probability that the particle reaches the origin for the
first time between $t$ and $t+dt$, the probability that the particle
has reached the origin at time $t$ and the variance $\langle
r^2\rangle$ of the distribution. We find that the reduced particle has
a normal diffusion behavior for small times with a gravity-modified
diffusion coefficient $\langle
r^2\rangle=r_{0}^2+(4k_B/\xi\mu)(T-T_{*})t$ and an anomalous diffusion
for large times $\langle r^2\rangle\propto t^{1-T_*/T}$. In
particular, the variance increases with time when $T>T_*$ and tends to
zero for $t\rightarrow +\infty$ when $T<T_*$. In Sec. \ref{sec_b}, we
consider the case of two self-gravitating Brownian particles in a
bounded domain and discuss the differences with the case of an
infinite domain. Finally, in Sec. \ref{sec_bp} we show that our study
also describes the large time asymptotics of the Smoluchowski-Poisson
system (or Keller-Segel model) for $T<T_c=GMm/(4k_B)$. Indeed, in the
post-collapse regime, the system is made of a growing central Dirac
peak (condensate) surrounded by a dilute halo whose dynamical
evolution is eventually described by a Fokker-Planck equation similar
to the one studied in the case of $N=2$ particles. We find that the
saturation of the mass of the condensate to the total mass is
algebraic in an infinite domain and exponential in a bounded domain
and we characterize it precisely. In Sec. \ref{sec_lfp}, we briefly
generalize our results to the logarithmic Fokker-Planck equation in
$d$ dimensions. 
The Appendices provide complements such as the
deterministic limit $T=0$ (Sec. \ref{sec_z}), the van Kampen
classification (Sec. \ref{sec_vk}), the correlation functions
(Sec. \ref{sec_corr}) and the general form of the virial theorem for
Brownian particles with power law interaction (Sec. \ref{sec_virg}).

We may note that our study bears some similarities with the
stochastic motion (induced by viscosity) of two point vortices studied
by Agullo \& Verga \cite{agullo}. However, there also exists crucial
differences between the two problems since in our case the interaction
is radial leading to the formation of Dirac peaks while in the case of
point vortices the interactions is rotational leading to the formation
of a spiral structure.

We may also note that the statistical mechanics of $N=2$ particles in
gravitational interaction has been considered by Padmanabhan
\cite{paddy} in $d=3$ (and generalized by Chavanis \cite{epjb} for the
dimensions $d=1$ and $d=2$) in the microcanonical and canonical
ensembles. However, these authors consider the {\it equilibrium}
statistical mechanics of $N=2$ self-gravitating particles in a box,
and with a small-scale cut-off, while we consider here the {\it
dynamical} evolution of $N=2$ self-gravitating Brownian particles in a
finite or infinite domain without small-scale cut-off. Therefore, we
address the time dependent problem and investigate the formation of
Dirac peaks.

Finally, the particular character of the dimension $d=2$ in gravity is
well-known. We refer for example to
\cite{salzberg,klb,paddy,paddy2,kiessling,at,ap,sc,epjb,marginal} for more details
and further references.

\section{The position of the problem}
\label{sec_position}

\subsection{The $N$-body problem}
\label{sec_nb}

We consider a system of $N$ overdamped Brownian particles with mass
$m_{\alpha}$ in gravitational interaction in a
space of dimension $d$. Their motion is described by the coupled
stochastic equations \cite{epjb}:
\begin{equation}
\frac{d{\bf r}_\alpha}{dt}=-\frac{1}{\xi m_\alpha}\nabla_\alpha U({\bf r}_1,...,{\bf r}_N)+\sqrt{2D_\alpha}{\bf B}_\alpha(t),  \label{nb1}
\end{equation}
with
\begin{equation}
U({\bf r}_1,...,{\bf r}_N)=-\frac{G}{d-2} \sum_{\alpha<\beta}\frac{m_\alpha m_\beta}{|{\bf r}_\alpha-{\bf r}_\beta|^{d-2}},  \label{nb2b}
\end{equation}
for $d\neq 2$ and
\begin{equation}
U({\bf r}_1,...,{\bf r}_N)=G \sum_{\alpha<\beta}m_\alpha m_\beta \ln |{\bf r}_\alpha-{\bf r}_\beta|,  \label{nb2}
\end{equation}
for $d=2$. Here, $\xi$ is the friction coefficient and ${\bf
B}_{\alpha}(t)$ is a white noise satisfying $\langle {\bf
B}_{\alpha}(t)\rangle=0$ and $\langle
B_{\alpha,i}(t)B_{\beta,j}(t')\rangle=\delta_{ij}\delta_{\alpha\beta}\delta(t-t')$
where $\alpha=1,...,N$ refers to the particles and $i=1,...,d$ to the
coordinates of space.  The diffusion coefficient is given by the
Einstein relation
\begin{equation}
D_\alpha=\frac{k_B T}{\xi m_\alpha}, \label{nb3}
\end{equation}
where $T$ is the temperature. We assume that the friction $\xi$ is the same for all the particles.

From these stochastic equations, it is possible to derive the
Fokker-Planck equation for the $N$-body distribution $P_{N}({\bf
r}_1,...,{\bf r}_{N},t)$ and then write the BBGKY-hierarchy for the
reduced distributions \cite{hb2,epjb}. Let us consider for brevity the
single-species system. The proper thermodynamic limit corresponds to
$N\rightarrow +\infty$ in such a way that the normalized temperature
$\eta=\beta GMm/R^{d-2}$ is of order unity. In that limit, it can be
shown that the mean field approximation becomes exact so that the
$N$-body distribution factorizes in a product of $N$ one-body
distributions: $P_{N}({\bf r}_1,...,{\bf
r}_{N},t)=\prod_{\alpha}P_{1}({\bf r}_{\alpha},t)$
\cite{hb2}. Furthermore, the one-body distribution, or equivalently the
smooth density field $\rho({\bf r},t)=NmP_{1}({\bf r},t)$, is solution of 
the Smoluchowski-Poisson system \cite{hb2}:
\begin{equation}
\frac{\partial\rho}{\partial t}=\nabla\cdot \left\lbrack \frac{1}{\xi}\left (\frac{k_{B}T}{m}\nabla\rho+\rho\nabla\Phi\right )\right\rbrack, \label{nb4}
\end{equation}
\begin{equation}
\Delta\Phi=S_d G\rho. \label{nb5}
\end{equation}
The equations generalizing Eqs. (\ref{nb4})-(\ref{nb5}) for the multi-species case are given in \cite{epjb,sopikmulti}.

Up to a change of notations, these equations are isomorphic to a
simplified version of the Keller-Segel model of chemotaxis that is
valid in the limit of large diffusivity of the chemical and in the
absence of degradation \cite{crrs}.

\subsection{The case $N=2$: the reduced particle}
\label{sec_r}

From now on, we consider only $N=2$ self-gravitating Brownian
particles in $d=2$. In that case, the stochastic equations
(\ref{nb1})-(\ref{nb2}) reduce to
\begin{equation}
\frac{d{\bf r}_1}{dt}=-\frac{Gm_2}{\xi}\frac{{\bf r}_1-{\bf r}_2}{|{\bf r}_1-{\bf r}_2|^2}+\sqrt{2D_1}{\bf B}_1(t),  \label{r1}
\end{equation}
\begin{equation}
\frac{d{\bf r}_2}{dt}=\frac{Gm_1}{\xi}\frac{{\bf r}_1-{\bf r}_2}{|{\bf r}_1-{\bf r}_2|^2}+\sqrt{2D_2}{\bf B}_2(t),  \label{r2}
\end{equation}
with $D_{1}=k_B T/\xi m_1$ and $D_{2}=k_B T/\xi m_2$.  Like for the
standard two-body problem, we introduce the center of mass
\begin{equation}
{\bf R}=\frac{m_1 {\bf r}_1+m_2 {\bf r}_2}{M}, \qquad M=m_1+m_2, \label{r3}
\end{equation}
and the reduced particle
\begin{equation}
{\bf r}={\bf r}_2-{\bf r}_1, \qquad \mu=\frac{m_1 m_2}{m_1+m_2}. \label{r4}
\end{equation}
Concerning the motion of the center of mass, we have
\begin{equation}
\frac{d{\bf R}}{dt}=\frac{1}{M}(m_1\sqrt{2D_1}{\bf B}_1(t)+m_2\sqrt{2D_2}{\bf B}_2(t))\equiv {\bf Q}(t), \label{r5}
\end{equation}
where the noise satisfies
\begin{equation}
\langle Q_i(t)Q_j(t')\rangle=\frac{2k_B T}{M\xi}\delta_{ij}\delta(t-t'). \label{r6}
\end{equation}
Therefore, the center of mass undergoes a pure Brownian motion of the form
\begin{equation}
\frac{d{\bf R}}{dt}=\sqrt{2D_*}{\bf B}(t),  \label{r7}
\end{equation}
with a diffusion coefficient
\begin{equation}
D_*=\frac{k_B T}{\xi M}. \label{r8}
\end{equation}
Concerning the motion of the reduced particle, we have
\begin{eqnarray}
\frac{d{\bf r}}{dt}+\frac{G}{\xi}(m_1+m_2)\frac{\bf r}{r^2}\qquad\qquad\nonumber\\
=\sqrt{2D_2}{\bf B}_2(t)
-\sqrt{2D_1}{\bf B}_1(t)\equiv {\bf S}(t), \label{r9}
\end{eqnarray}
where the noise satisfies
\begin{equation}
\langle S_i(t)S_j(t')\rangle=\frac{2k_B T}{\mu\xi}\delta_{ij}\delta(t-t'). \label{r10}
\end{equation}
Therefore, the reduced particle undergoes a Brownian motion in a central potential of the form
\begin{equation}
\frac{d{\bf r}}{dt}=-\frac{Gm_1m_2}{\xi\mu}\frac{\bf r}{r^2}+\sqrt{2D}{\bf B}(t),  \label{r11}
\end{equation}
with a diffusion coefficient
\begin{equation}
D=\frac{k_B T}{\xi \mu}. \label{r12}
\end{equation}

\subsection{The naive virial theorem}
\label{sec_nv}

Let us introduce the total moment of inertia
\begin{equation}
I_{tot}(t)=m_1\langle r_1^2\rangle+m_2\langle r_2^2\rangle. \label{nv1}
\end{equation}
In \cite{epjb} (see also Appendix \ref{sec_virg}), an exact closed expression  of the virial theorem valid for an arbitrary number of self-gravitating Brownian particles in $d=2$ has been obtained. For $N=2$, it writes 
\begin{equation}
\frac{1}{4}\xi\dot I_{tot}=2k_B(T-T_c),  \label{nv2}
\end{equation}
with the critical temperature
\begin{equation}
k_B T_c=\frac{G m_1m_2}{4}.  \label{nv3}
\end{equation}
It is instructive to recover this result in a different manner. The
positions of the particles $1$ and $2$ can be expressed in terms of
${\bf r}$ (reduced particle) and ${\bf R}$ (center of mass) as
\begin{equation}
{\bf r}_1=\frac{M{\bf R}-m_2{\bf r}}{M},\qquad  {\bf r}_2=\frac{M{\bf R}+m_1{\bf r}}{M}.
\label{nv4}
\end{equation}
Substituting these relations in Eq. (\ref{nv1}), we obtain after
straightforward algebra
\begin{equation}
I_{tot}(t)=M\langle R^2\rangle+\mu\langle r^2\rangle,  \label{nv5}
\end{equation}
a relation which was of course expected. Now, the Fokker-Planck equation associated with the stochastic motion  (\ref{r11}) of
the reduced particle is
\begin{equation}
\xi{\partial P\over \partial t}=\nabla\cdot \left (\frac{k_B T}{\mu}\nabla P+P\frac{Gm_1m_2}{\mu}\frac{{\bf r}}{r^2}\right ).\label{nv6}
\end{equation}
Taking the time derivative of
\begin{equation}
\langle r^2\rangle=\int P r^2\, d{\bf r},\label{nv7}
\end{equation}
and using simple integrations by parts, we naively\footnote{We shall see later that this expression is in fact incorrect.} obtain
\begin{equation}
\frac{1}{4}\xi\mu\frac{d\langle r^2\rangle}{dt}=k_B T-\frac{Gm_1m_2}{2}.  \label{nv8}
\end{equation}
This relation exhibits a critical temperature
\begin{equation}
k_B T_*=\frac{G m_1m_2}{2}.  \label{nv9}
\end{equation}
Introducing the moment of inertia of the reduced particle $I(t)=\mu\langle r^2\rangle$, we can rewrite Eq. (\ref{nv8}) as
\begin{equation}
\frac{1}{4}\xi\frac{dI}{dt}=k_B (T-T_*).  \label{nv10}
\end{equation}
The mean square displacement of  the reduced particle satisfies
\begin{equation}
\langle r^2\rangle=\frac{4k_B}{\xi\mu}(T-T_*)t+\langle r^2\rangle_0. \label{nv11}
\end{equation}
This is a normal diffusion with a gravity modified diffusion coefficient
\begin{equation}
D(T)=\frac{k_B T}{\xi\mu}\left (1-\frac{T_*}{T}\right ).
\label{ew}
\end{equation}
The variance increases for $T>T_*$ and
tends to zero in a finite time for $T<T_*$. On the other hand, the
Fokker-Planck equation associated to the stochastic motion (\ref{r7}) of the
center of mass is simply
\begin{equation}
\xi{\partial P\over \partial t}=\frac{k_B T}{M}\Delta P, \label{nv12}
\end{equation}
and we classically obtain the relation
\begin{equation}
\frac{1}{4}\xi M\frac{d\langle R^2\rangle}{dt}=k_B T.  \label{nv13}
\end{equation}
Finally, summing Eqs. (\ref{nv8}) and (\ref{nv13}) and using
Eq. (\ref{nv5}), we recover Eq. (\ref{nv2}). We now clearly see the
origin of the two temperatures $T_c$ and $T_*=2T_c$ that were reported
in \cite{epjb}. In the case $N=2$, the critical temperature $T_*$ is
associated to the dynamics of the reduced particle while the critical
temperature $T_c$ enters in the expression of virial theorem for the
total moment of inertia (reduced particle and center of mass). This
distinction is further discussed in Appendix \ref{sec_mict} in the
general case of $N$ particles.

\subsection{The problem}
\label{sec_prob}

In fact, there is a flaw in the above derivation of the virial theorem
because we have naively assumed that the normalization $\int P({\bf r},t)\, d{\bf r}=1$ is conserved in
time. However, as we shall see, this is not correct. The normalization
is not conserved in time because the reduced particle can reach the
origin $r=0$ and be ``lost'' by the system (if it reaches the origin,
it remains there for ever). This corresponds to the coalescence of the
two particles, resulting in the formation of a Dirac peak, i.e. a 
new particle of mass $m_1+m_2$. As a result of these
``trapping'' events
\begin{equation}
\int P({\bf r},t)\, d{\bf r}\neq 1, \label{mf1}
\end{equation}
and we must reconsider the problem in more detail.

\section{Brownian particle in a Newtonian potential in two dimensions}
\label{sec_bn}

\subsection{The Fokker-Planck equation}
\label{sec_fp}

Let $P({\bf r},t)$ denote the  probability density of finding the  reduced
particle in ${\bf r}$ at time $t$. The evolution of $P({\bf r},t)$
is governed by the Fokker-Planck equation\footnote{Note that, for $d=1$, the Fokker-Planck equation for the reduced particle corresponds to a V-shaped potential $U(x)=Gm_1m_2|x|$ that relaxes to $P_s(x)=\frac{1}{2}\beta Gm_1m_2 e^{\beta Gm_1m_2|x|}$ \cite{risken}.}
\begin{equation}
\xi{\partial P\over \partial t}=\nabla\cdot \left (\frac{k_B T}{\mu}\nabla P+P\frac{Gm_1m_2}{\mu}\frac{{\bf r}}{r^2}\right ). \label{fp1}
\end{equation}
The initial
distribution $P_0({\bf r})$ is normalized such that $\int
P_{0}({\bf r})d{\bf r}=1$. Introducing
\begin{equation}
D=\frac{k_B T}{\mu\xi},\quad \beta=\frac{1}{k_B T}, \quad U=Gm_1m_2\ln r,   \label{fp2}
\end{equation}
the Fokker-Planck equation can be rewritten
\begin{equation}
{\partial P\over \partial t}=\nabla\cdot \left\lbrack D\left (\nabla P+\beta P\nabla U\right )\right\rbrack.   \label{fp3}
\end{equation}
In the absence of small and large scale cut-offs, this equation has no steady state since the distribution $P=A/r^{\beta G m_1 m_2}$ is not normalizable. We assume that the initial distribution $P_0({\bf r})$ is radially
symmetric, so that $P({\bf r},t)$ is radially symmetric for all times.
Therefore, we can write the Fokker-Planck equation in the form
\begin{equation}
{\partial P\over \partial t}={1\over r}{\partial\over\partial r}\biggl\lbrace D r\biggl ({\partial P\over\partial r}+P \frac{\beta G m_1m_2}{r}\biggr )\biggr\rbrace.     \label{fp4}
\end{equation}
As discussed previously, the probability is not conserved because the
reduced particle may reach the origin $r=0$ and form a Dirac peak (the
two particles coalesce).  The probability that the particle has not
reached $r=0$ at time $t$ is
\begin{equation}
\chi(t)=\int_0^{+\infty}P(r,t)2\pi r\, dr.   \label{fp5}
\end{equation}
Taking the time derivative of this quantity and using the Fokker-Planck equation (\ref{fp4}) we obtain
\begin{equation}
\dot \chi(t)=-2\pi D\beta Gm_1m_2 P(0,t),   \label{fp6}
\end{equation}
which is non zero since $P(0,t)\neq 0$. Therefore, the probability for the particle to form a Dirac peak between $t$ and $t+dt$ (i.e. to reach $r=0$ for the first time between $t$ and $t+dt$) is
\begin{equation}
\dot \chi_D(t)=2\pi D\beta Gm_1m_2 P(0,t),  \label{fp7}
\end{equation}
and the probability for the particle to have formed a Dirac peak at time $t$ (i.e. to have reached $r=0$ at time $t$) is
\begin{equation}
\chi_D(t)=2\pi D\beta Gm_1m_2\int_{0}^{t} P(0,\tau)\, d\tau.  \label{fp8}
\end{equation}
We obviously have $\chi_D(t)=1-\chi(t)$. We can now obtain the proper
form of the virial theorem associated to the Fokker-Planck equation
(\ref{fp1}). Introducing the moment of inertia of the reduced particle
\begin{equation}
I(t)=\int P \mu r^2\, d{\bf r},
\label{fp9}
\end{equation}
we easily obtain the virial theorem
\begin{equation}
\frac{1}{4}\xi \frac{dI}{dt}=\chi(t)k_B (T-T_*),
\label{fp10}
\end{equation}
instead of Eq. (\ref{nv10}). It has to be noted that this relation is not closed since it depends on $\chi(t)$ that must be obtained by solving the Fokker-Planck equation (\ref{fp1}).

\subsection{The associated Schr\"odinger equation}
\label{sec_s}

Let us consider a general Fokker-Planck equation of the form
\begin{equation}
{\partial P\over \partial t}={\partial \over\partial {\bf r}}\cdot \biggl\lbrack D\biggl ({\partial P\over\partial {\bf r}}+\beta P{\partial U\over\partial {\bf r}}\biggr )\biggr\rbrack.   \label{s1}
\end{equation}
For a spherically symmetric distribution in $d$ dimensions, it can be
rewritten
\begin{equation}
{\partial P\over \partial t}={1\over r^{d-1}}{\partial\over\partial r}\biggl\lbrace D(r)r^{d-1}\biggl ({\partial P\over\partial r}+\beta P{\partial U\over\partial r}\biggr )\biggr\rbrace.     \label{s2}
\end{equation}
As is well-known \cite{risken}, we can transform this Fokker-Planck equation into a Schr\"odinger equation (with imaginary time) by setting
\begin{equation}
P({\bf r},t)=e^{-{1\over 2}\beta U({\bf r})}\psi({\bf r},t).   \label{s3}
\end{equation}
This yields
\begin{eqnarray}
{\partial\psi\over\partial t}={\partial\over\partial {\bf r}}\cdot \biggl (D{\partial\psi\over\partial {\bf r}}\biggr )+V({\bf r})\psi=(H+V)\psi, \label{s4}
\end{eqnarray}
with the potential
\begin{eqnarray}
V({\bf r})= {1\over 2}\beta  {\partial\over\partial {\bf r}}\cdot \biggl (D{\partial U\over\partial {\bf r}}\biggr )-{1\over 4}D\beta^{2}\biggl ({\partial U\over\partial {\bf r}}\biggr )^{2}. \label{s5}
\end{eqnarray}
For a spherically symmetric distribution,  the Schr\"odinger equation (\ref{s4}) can be rewritten
\begin{eqnarray}
{\partial\psi\over\partial t}=\frac{1}{r^{d-1}}\frac{\partial}{\partial r}\left (r^{d-1}D(r)\frac{\partial\psi}{\partial r}\right )+V(r)\psi,
 \label{s6}
\end{eqnarray}
with
\begin{eqnarray}
V(r)=\frac{1}{2}\beta \frac{1}{r^{d-1}}\frac{d}{d r}\left (r^{d-1}D(r)\frac{dU}{dr}\right )
-\frac{1}{4}\beta^2 D(r)\left (\frac{dU}{dr}\right )^{2}.\nonumber\\
 \label{s7}
\end{eqnarray}
Making the separation of variables
\begin{eqnarray}
\psi=e^{-\lambda t}\phi(r), \label{s8}
\end{eqnarray}
we obtain the eigenvalue equation
\begin{eqnarray}
{1\over r^{d-1}}{d\over d r}\biggl\lbrack r^{d-1}D(r){d\phi\over d r}\biggr \rbrack+(V(r)+\lambda)\phi=0.\label{s9}
\end{eqnarray}
Let us note  $\lambda_n$ the eigenvalues and $\phi_n$ the corresponding eigenfunctions. The eigenfunctions are orthogonal with respect to the scalar product
\begin{eqnarray}
\langle f |g\rangle=\int f(r)g(r)S_{d}r^{d-1}dr.\label{s10}
\end{eqnarray}
We also normalize them so that $\langle
\phi_n|\phi_m\rangle=\delta_{nm}$. Then, any function can be expanded in the form
\begin{eqnarray}
h(r)=\sum_{n=1}^{+\infty}\langle h |\phi_{n}\rangle \phi_{n}(r).
\label{s11}
\end{eqnarray}
If the spectrum is continuous, the sum over $n$ must be replaced by an integral over
$\lambda\ge 0$.

\subsection{The general solution}
\label{sec_gs}

We consider the Green function $P(r,t|r_{0})$ which corresponds to the initial condition
\begin{eqnarray}
P(r,0|r_0)={\delta(r-r_{0})\over S_{d}r_{0}^{d-1}}. \label{gs1}
\end{eqnarray}
The solution on the Fokker-Planck equation (\ref{s2}) can be expanded on the eigenfunctions in the form
\begin{equation}
P(r,t|r_0)=\sum_{n=1}^{+\infty}A_{n}e^{-\lambda_{n}t}e^{-{1\over 2}\beta U(r)}\phi_{n}(r).     \label{gs2}
\end{equation}
Noting that
\begin{equation}
{\delta(r-r_{0})\over S_{d}r_{0}^{d-1}}=\sum_{n=1}^{+\infty}\phi_n(r_{0})\phi_{n}(r),  \label{gs3}
\end{equation}
and using the initial condition (\ref{gs1}), we finally obtain
\begin{equation}
P(r,t|r_0)={e^{-{1\over 2}\beta (U(r)-U(r_{0}))}}\sum_{n=1}^{+\infty} e^{-\lambda_{n}t}\phi_{n}(r_{0})\phi_{n}(r).     \label{gs4}
\end{equation}

\subsection{The case of a logarithmic potential in $d=2$}
\label{sec_lp}

\begin{figure}
\begin{center}
\includegraphics[clip,scale=0.3]{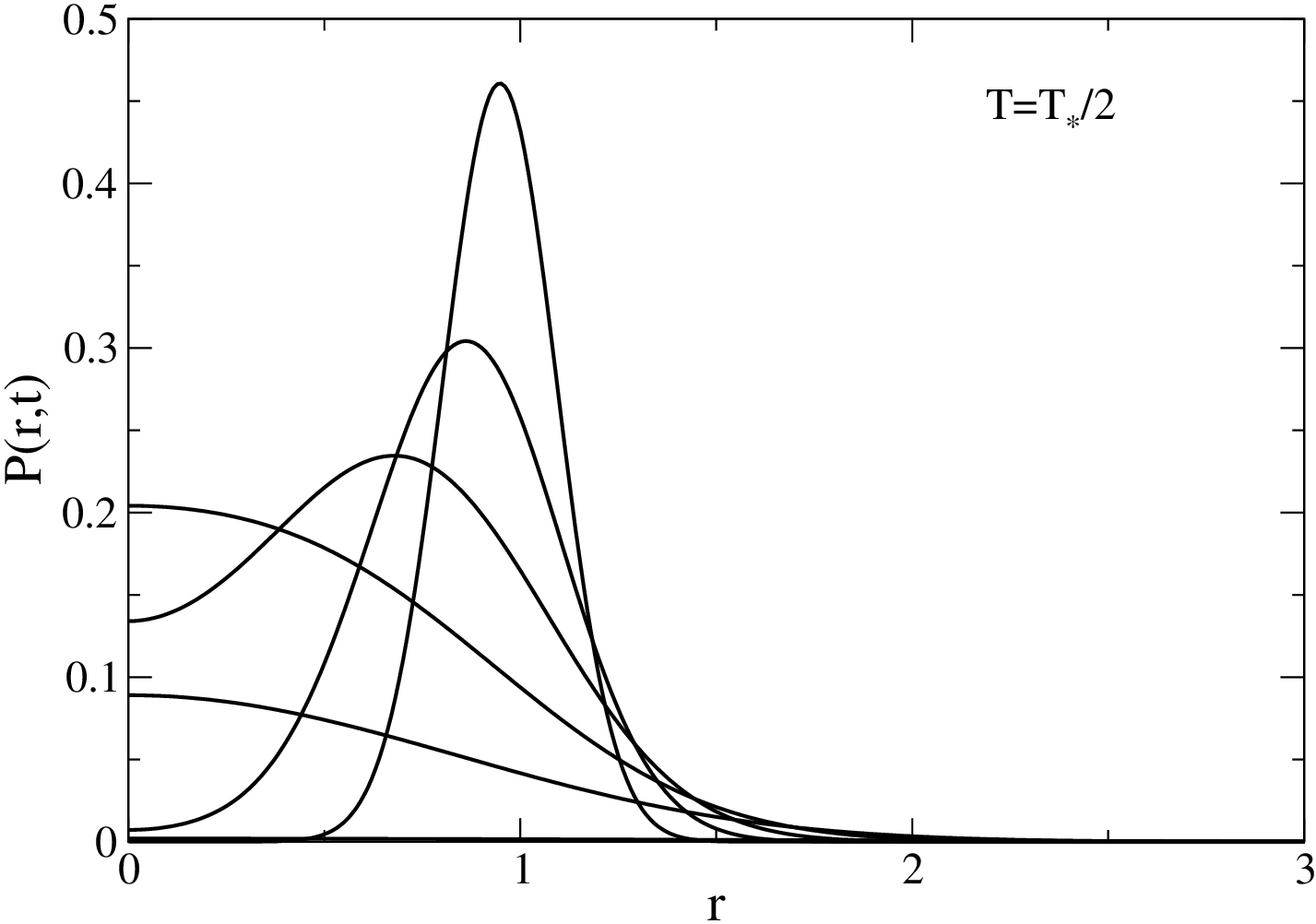}
\caption{Probability density  of finding the particles at a distance ${\bf r}$ from each other, at different times $t$ and for a temperature $T=T_*/2$ ($a=2$). From top to bottom: $t=0.01, 0.025, 0.05, 0.1, 0.2, 1$. }
\label{prta2}
\end{center}
\end{figure}

For the Fokker-Planck equation (\ref{fp4}), we have $d=2$,
$D=k_BT/(\xi\mu)$ and $U=Gm_1m_2\ln r$. The potential $V(r)$ arising
in the corresponding Schr\"odinger equation (\ref{s6}) is
\begin{equation}
V(r)=-D\left (\frac{\beta Gm_1m_2}{2r}\right )^2.
\label{lp1}
\end{equation}
Therefore, if we assume that initially
\begin{eqnarray}
P(r,0)={\delta(r-r_{0})\over 2\pi r_{0}},  \label{lp2}
\end{eqnarray}
the solution of  the Fokker-Planck equation (\ref{fp4}) can be written
\begin{equation}
P(r,t)=\left (\frac{r_0}{r}\right )^{\frac{1}{2}\beta Gm_1m_2}\int_{0}^{+\infty}e^{-\lambda t}\phi_{\lambda}(r_{0})\phi_{\lambda}(r)\, d\lambda,
\label{lp3}
\end{equation}
where $\phi_{\lambda}(r)$ is solution of the differential equation
\begin{equation}
r^2\phi''+r\phi'+\left (\frac{\lambda}{D} r^2-a^2\right )\phi=0,
\label{lp4}
\end{equation}
where
\begin{equation}
a=\frac{\beta Gm_1m_2}{2}=\frac{T_*}{T}.
\label{lp5}
\end{equation}
Equation (\ref{lp4}) is a Bessel differential equation that can be solved analytically. The solutions that are finite at the origin are of the form
\begin{equation}
\phi_\lambda(r)= J_{a}(\sqrt{\lambda/D}r).
\label{lp6}
\end{equation}
Substituting Eq. (\ref{lp6}) in Eq. (\ref{lp3}), the solution of the Fokker-Planck equation (\ref{fp4}) can
be written
\begin{eqnarray}
P(r,t)=2D\left (\frac{r_0}{r}\right )^{a}A^2\int_{0}^{+\infty}e^{-D\lambda^2 t}J_{a}(\lambda r_{0})J_{a}(\lambda r)\, \lambda d\lambda,\nonumber\\
\label{lp7}
\end{eqnarray}
where we have made the change of notation $\lambda\rightarrow
D\lambda^2$ for convenience. The normalization
constant $A$ is determined so as to recover the initial condition
(\ref{lp2}) as $t\rightarrow 0$. Taking $t=0$ in Eq. (\ref{lp7}), we get
\begin{eqnarray}
P(r,0)=2D\left (\frac{r_0}{r}\right )^{a}A^2\int_{0}^{+\infty}J_{a}(\lambda r_{0})J_{a}(\lambda r)\, \lambda d\lambda.\nonumber\\
\label{lp7b}
\end{eqnarray}
Using the closure relation
\begin{eqnarray}
\int_{0}^{+\infty}J_{a}(ux)J_{a}(vx)x\, dx=\frac{1}{u}\delta(u-v),
\label{lp7cbis}
\end{eqnarray}
we obtain
\begin{eqnarray}
P(r,0)=\frac{2DA^2}{r_0}\delta(r-r_0).
\label{lp7c}
\end{eqnarray}
Comparing with Eq. (\ref{lp2}), we find that $A^2=1/(4\pi D)$. Therefore, the solution of the Fokker-Planck equation (\ref{fp4}) with the initial condition (\ref{lp2}) is
\begin{eqnarray}
P(r,t)=\frac{1}{2\pi}\left (\frac{r_0}{r}\right )^{a}\int_{0}^{+\infty}e^{-D\lambda^2 t}J_{a}(\lambda r_{0})J_{a}(\lambda r)\, \lambda d\lambda.\nonumber\\
\label{lp7d}
\end{eqnarray}
Using the identity
\begin{eqnarray}
\int_{0}^{+\infty}e^{-\rho^2 x^2}J_{\gamma}(\alpha x)J_{\gamma}(\beta x)\, x dx=\frac{1}{2\rho^2}e^{-\frac{\alpha^2+\beta^2}{4\rho^2}}I_{\gamma}\left (\frac{\alpha\beta}{2\rho^2}\right ),\nonumber\\
\label{lp8}
\end{eqnarray}
valid for $\gamma>-1$, we find that it can finally be written
\begin{equation}
P(r,t)=\left (\frac{r_0}{r}\right )^{a}\frac{1}{4\pi Dt}e^{-\frac{r_0^2+r^2}{4Dt}} I_{a}\left (\frac{rr_0}{2Dt}\right ).
\label{lp12}
\end{equation}
The distribution $P(r,t)$ is plotted in Fig. \ref{prta2} at different
times and for $T/T_*=1/2$ (corresponding to $a=2$).

Using the identity
\begin{equation}
I_{a}(x)\sim \frac{e^{x}}{\sqrt{2\pi x}}, \qquad (x\rightarrow +\infty)
\label{lp10}
\end{equation}
we get for $t\rightarrow 0$:
\begin{equation}
P(r,t)\sim \frac{1}{4\pi r_0\sqrt{\pi D t}}e^{-\frac{(r-r_0)^2}{4Dt}},
\label{lp11}
\end{equation}
which tends to Eq. (\ref{lp2}) as expected. On the other hand, for
$t\rightarrow +\infty$, the probability tends to zero meaning that the
particle has been absorbed in $r=0$ after a sufficiently long time so
that it is ultimately lost by the system. For $r\rightarrow +\infty$, using the identity (\ref{lp10}), we have
\begin{equation}
P(r,t)\sim\left (\frac{r_0}{r}\right )^{a}\frac{1}{4\pi}\frac{1}{\sqrt{\pi r r_0 D t}}e^{-\frac{r^2}{4Dt}}.
\label{lp13}
\end{equation}
For $r=0$, using the identity
\begin{equation}
I_{a}(x)\sim \frac{1}{\Gamma(a+1)}\left (\frac{x}{2}\right )^{a}, \qquad (x\rightarrow 0)
\label{lp14}
\end{equation}
we get
\begin{equation}
P(0,t)=\left (\frac{r_0}{2}\right )^{2a}\frac{1}{4\pi \Gamma(a+1)}\frac{1}{(Dt)^{1+a}}e^{-\frac{r_0^2}{4Dt}}.
\label{lp15}
\end{equation}
Finally, for $\beta=0$, the probability density (\ref{lp12}) becomes
\begin{equation}
P(r,t)=\frac{1}{4\pi D t}e^{-\frac{r_0^2+r^2}{4Dt}} I_{0}\left (\frac{rr_0}{2Dt}\right ),
\label{lp16}
\end{equation}
which is the solution of the diffusion equation in $d=2$. Indeed, in
the limit of infinite temperature, the gravity is negligible with
respect to diffusion.

\subsection{The probability to form a Dirac peak}
\label{sec_d}

\begin{figure}
\begin{center}
\includegraphics[clip,scale=0.3]{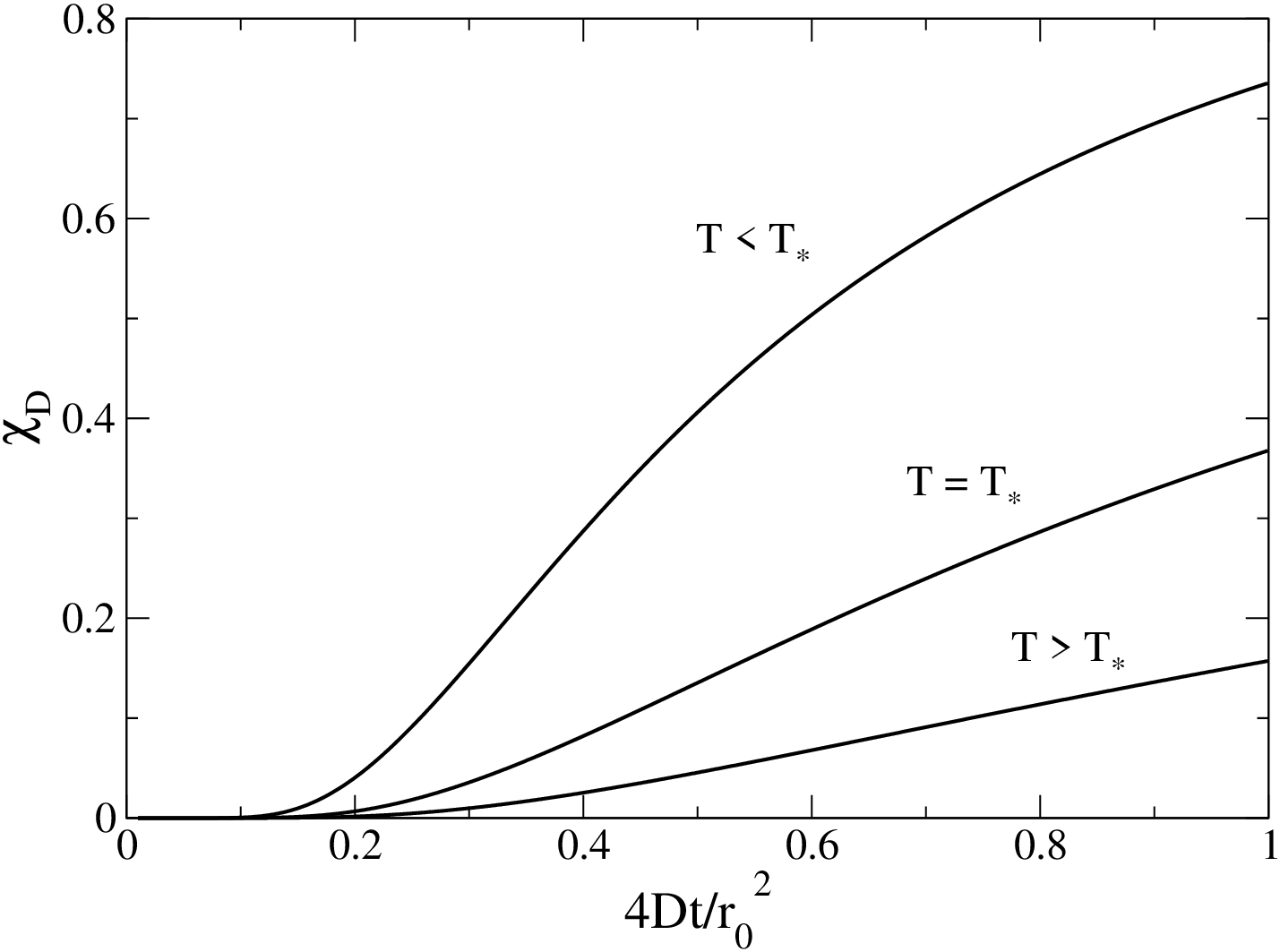}
\caption{Evolution of the function $\chi_D(t)$, giving the probability that the particle has formed a Dirac peak at time $t$, for different values of the temperature (we have taken $T/T_*=1/2$, $T/T_*=1$ and $T/T_*=2$). The probability increases more rapidly at smaller temperatures. }
\label{chiDpetitt}
\end{center}
\end{figure}

\begin{figure}
\begin{center}
\includegraphics[clip,scale=0.3]{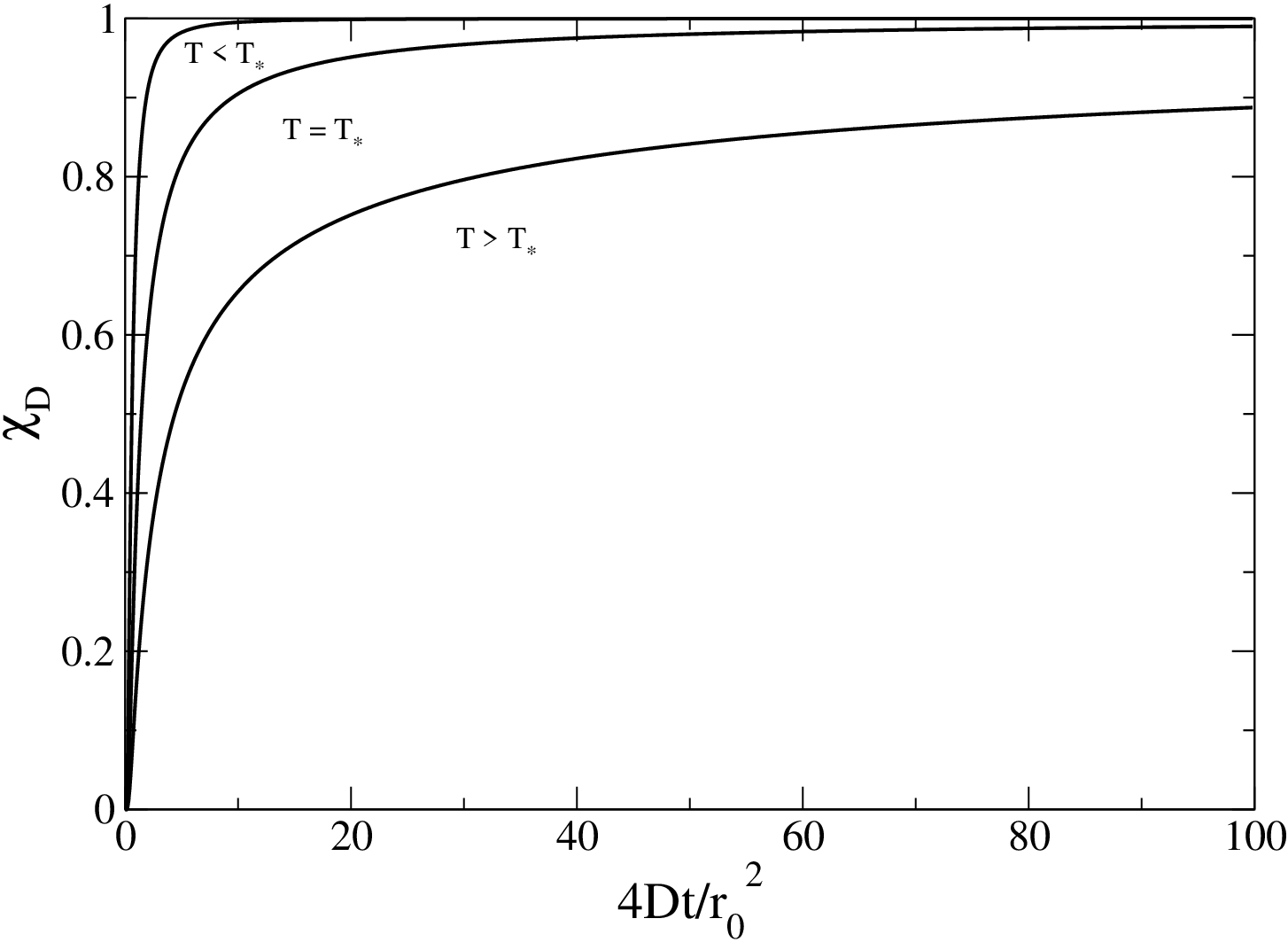}
\caption{Same as Fig. \ref{chiDpetitt} for larger times. }
\label{chiDgrandt}
\end{center}
\end{figure}

The probability that the particle has not reached $r=0$ at time $t$ is given by Eq. (\ref{fp5}). For the distribution (\ref{lp12}), the integral can be performed analytically and we obtain
\begin{equation}
\chi(t)=1-\frac{\Gamma_a\left (\frac{r_{0}^2}{4Dt}\right )}{\Gamma(a)},
\label{d1}
\end{equation}
where
\begin{equation}
\Gamma_a(x)=\int_{x}^{+\infty}t^{a-1}e^{-t}\, dt,
\label{d2}
\end{equation}
is the incomplete Gamma function. The probability decays because, as
time goes on, the particle has more and more chance to reach $r=0$ and form a
Dirac.  The probability that the particle reaches $r=0$ for the first
time between $t$ and $t+dt$ is given by Eq. (\ref{fp7}). Combining this
relation with Eq. (\ref{lp15}), we obtain
\begin{equation}
\dot \chi_{D}=D\left (\frac{r_0}{2}\right )^{2a}\frac{1}{\Gamma(a)}\frac{1}{(Dt)^{1+a}}e^{-\frac{r_0^2}{4Dt}}.
\label{d3}
\end{equation}
Integrating Eq. (\ref{d3}), we obtain the probability that the particle has formed a Dirac peak at time t:
\begin{equation}
\chi_{D}(t)=\frac{\Gamma_a\left (\frac{r_{0}^2}{4Dt}\right )}{\Gamma(a)},
\label{d4}
\end{equation}
and we check that $\chi(t)=1-\chi_{D}(t)$ as expected. The evolution
of $\chi_D(t)$ for different values of the temperature is shown in
Figs.  \ref{chiDpetitt} and \ref{chiDgrandt}.

For $t\rightarrow 0$, using the
expansion
\begin{eqnarray}
\Gamma_a(x)\sim x^{a-1}e^{-x}, \qquad (x\rightarrow +\infty),
\label{d7}
\end{eqnarray}
we obtain
\begin{equation}
\chi_{D}(t)\sim \frac{1}{\Gamma(a)}\left (\frac{r_{0}^{2}}{4Dt}\right )^{a-1}e^{-\frac{r_{0}^{2}}{4Dt}}, \qquad (t\rightarrow 0).
\label{d8}
\end{equation}
We see that the probability $\chi_{D}(t)\rightarrow 0$ for
$t\rightarrow 0$ due to the exponential factor. This tendency is reinforced by the algebraic factor for $T>T_*$
($a<1$) while it is reduced for $T<T_*$ ($a>1$).

For $t\rightarrow +\infty$, using the expansion
\begin{eqnarray}
\Gamma_a(x)\simeq \Gamma(a)-\frac{x^a}{a}, \qquad (x\rightarrow 0),
\label{d5}
\end{eqnarray}
we obtain
\begin{equation}
\chi_{D}(t)=1-\frac{1}{a\Gamma(a)}\left (\frac{r_{0}^{2}}{4Dt}\right )^a, \qquad (t\rightarrow +\infty).
\label{d6}
\end{equation}
Therefore, the probability that the particle has not formed a Dirac at
time $t$ decreases {\it algebraically} as $t^{-a}$.  Equation
(\ref{d6}) can be written in the form
\begin{equation}
\chi_{D}(t)=1-\left (\frac{t_{*}(a)}{t}\right )^a,
\label{d6a}
\end{equation}
where the time
\begin{equation}
t_{*}(a)=\frac{r_{0}^2}{4D}\frac{1}{\left\lbrack a\Gamma(a)\right\rbrack^{1/a}},
\label{d6b}
\end{equation}
gives an idea of the rapidity at which the Dirac forms as a function
of the temperature $a=T_*/T$. The function $t_{*}(a)$ is represented
in Fig. \ref{tstar} and its asymptotic behaviors are given in Appendix
\ref{sec_asym}. We find that $4Dt_*(a)/r_0^2\simeq 1.78107...$ for $a=0$ and 
$4Dt_*(a)/r_0^2\simeq 2.7182818/a...$ for $a\rightarrow +\infty$. We note that
$t_{*}(a)$ tends to a finite value for $a\rightarrow 0$ while we know
that the system does not form a Dirac peak for $a=0$ (indeed
$1-(t_*(a)/t)^a\rightarrow 0$ for $a\rightarrow 0$). Therefore, the
physical interpretation of $t_{*}(a)$ should be considered with
care. Another measure of the effect of the temperature on the
formation of the Dirac is provided by the quantity
\begin{equation}
\chi_{D}(1)=\frac{\Gamma_a(1)}{\Gamma(a)},
\label{d4b}
\end{equation}
corresponding to $\chi_D(t)$ evaluated at $t=r_0^2/4D$. This function is represented
in Fig. \ref{chi1} and its asymptotic behaviors are given in Appendix
\ref{sec_asym}. We find that $\chi_D(1)\sim 0.219384a$ for $a\rightarrow 0$
and $1-\chi_D(1)\sim \frac{1}{\sqrt{2\pi}}\frac{e^{a-1}}{a^{a+1/2}}$ for $a\rightarrow +\infty$.

\begin{figure}
\begin{center}
\includegraphics[clip,scale=0.3]{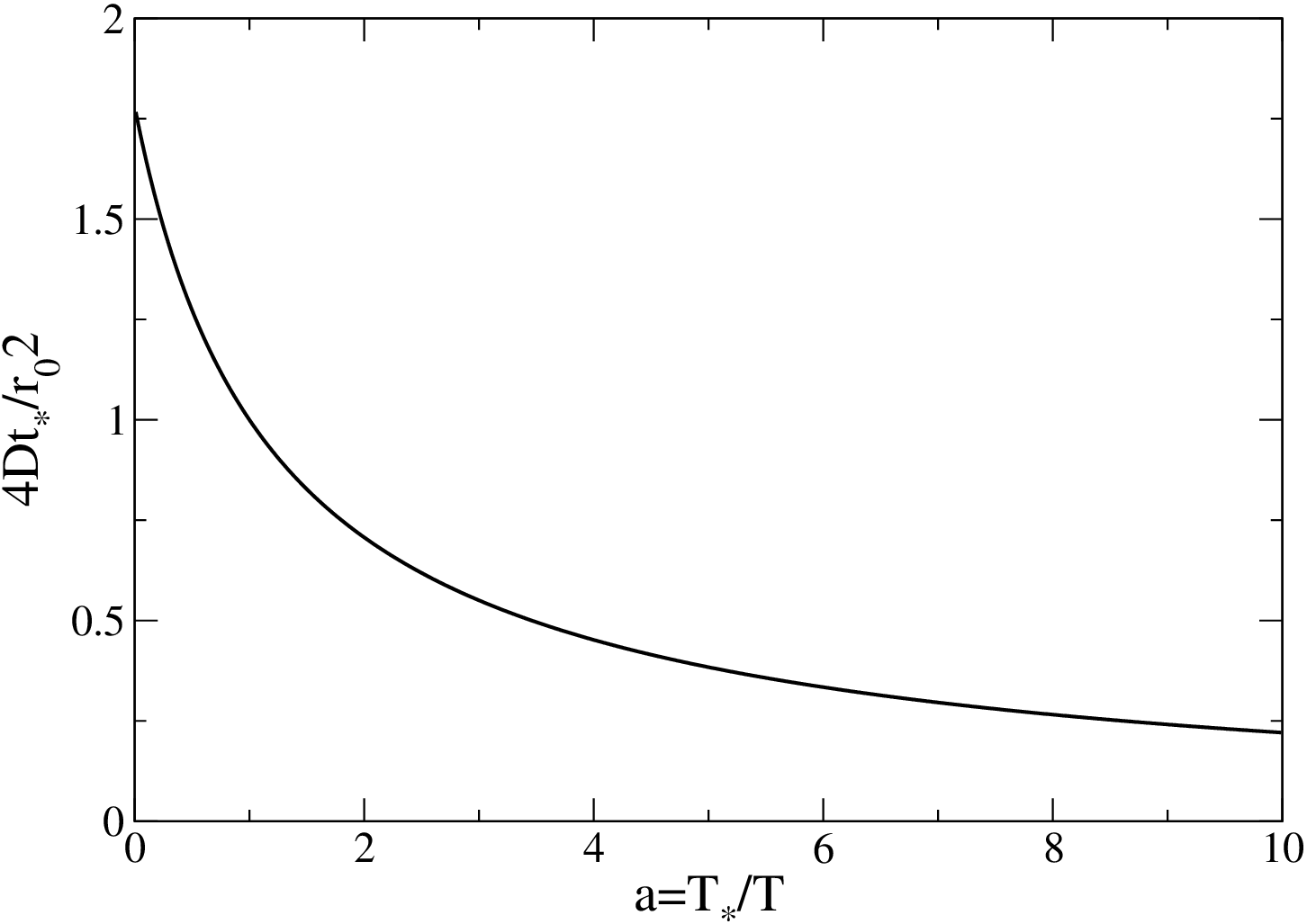}
\caption{Evolution of $t_{*}(a)$ as a function of the temperature $a=T_*/T$.}
\label{tstar}
\end{center}
\end{figure}

\begin{figure}
\begin{center}
\includegraphics[clip,scale=0.3]{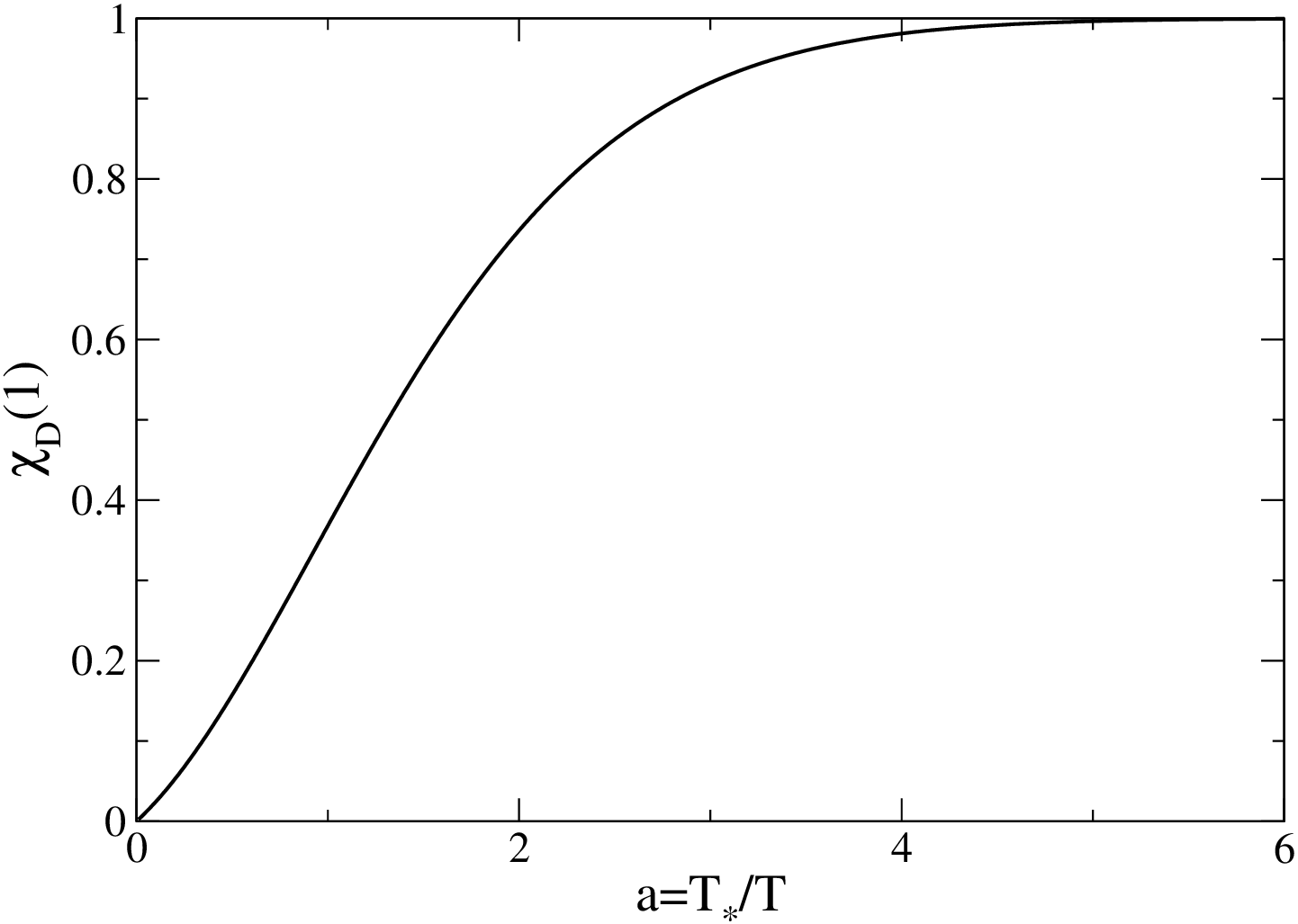}
\caption{Evolution of $\chi_D(1)$ as a function of the temperature $a=T_*/T$.}
\label{chi1}
\end{center}
\end{figure}

Finally, the normalized probability density can be written
\begin{equation}
P_{tot}({\bf r},t)=P({\bf r},t)+\chi_D(t)\delta({\bf r}),
\label{npd}
\end{equation}
where $P({\bf r},t)$ is given by Eq. (\ref{lp12}) and $\chi_D(t)$ by Eq. (\ref{d4}). We readily check that $\int P_{tot}({\bf r},t)\, d{\bf r}=1$.

\subsection{The moment of inertia}
\label{sec_m}

\begin{figure}
\begin{center}
\includegraphics[clip,scale=0.3]{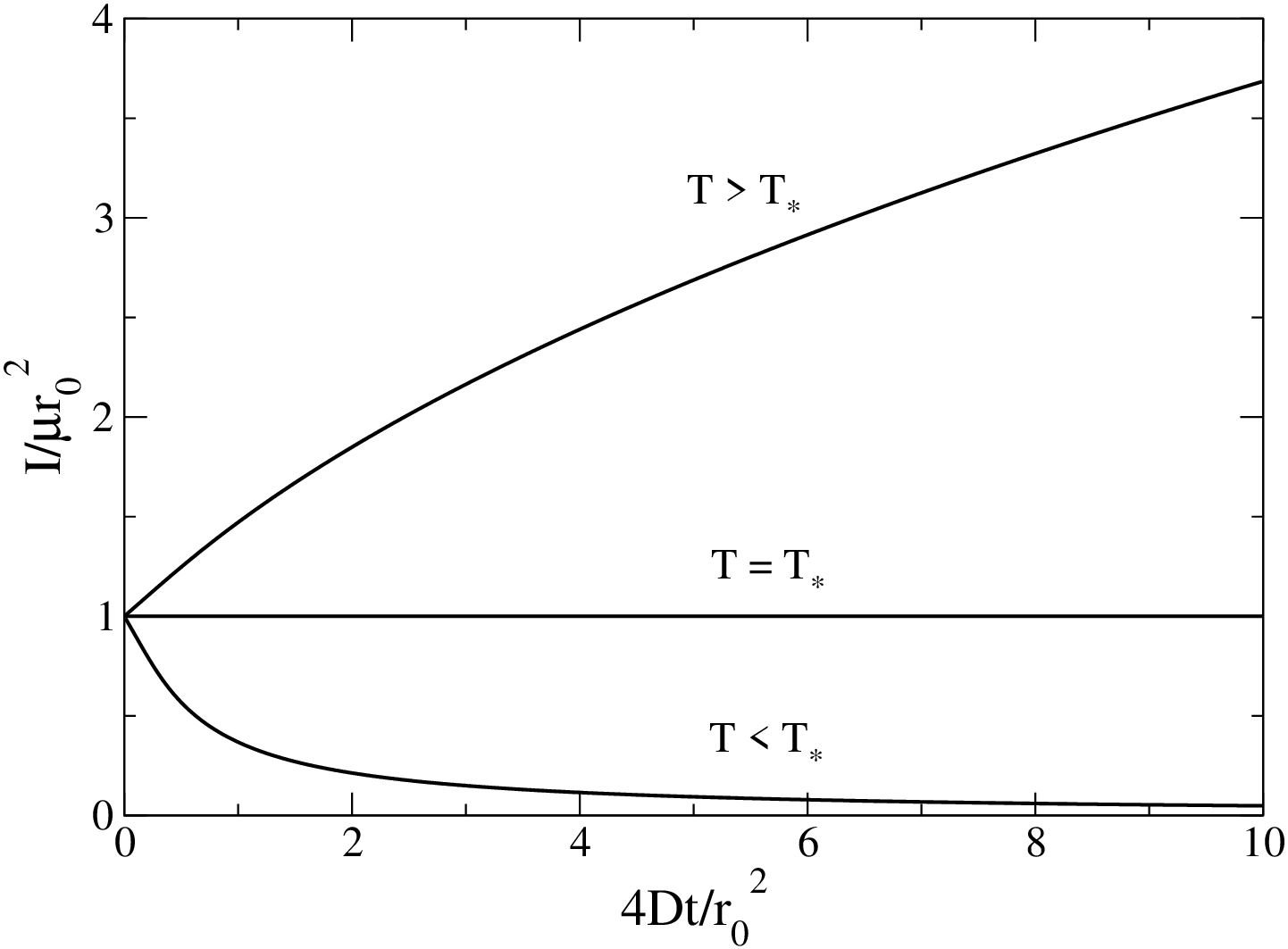}
\caption{Time evolution of the moment of inertia for different values of the temperature (we have taken $T/T_*=1/2$, $T/T_*=1$ and $T/T_*=2$). }
\label{i}
\end{center}
\end{figure}

The moment of inertia of the reduced particle is defined by
\begin{equation}
I(t)=\int_{0}^{+\infty}P(r,t)\mu r^2 2\pi r\, dr,
\label{m1}
\end{equation}
and the variance of the distribution (mean square displacement) is
\begin{equation}
\langle r^2\rangle=\frac{I(t)}{\mu}.
\label{m2}
\end{equation}
For the density distribution given by Eq. (\ref{lp12}), the integral can be calculated explicitly yielding
\begin{eqnarray}
I(t)=\frac{4\mu}{\Gamma(a)}\left (\frac{r_{0}^{2}}{4Dt}\right )^{a}Dt e^{-\frac{r_{0}^2}{4Dt}}\nonumber\\
+\mu(r_{0}^{2}+4D(1-a)t)\left\lbrack 1-\frac{\Gamma_{a}\left (\frac{r_0^2}{4Dt}\right )}{\Gamma(a)}\right\rbrack.
\label{m3}
\end{eqnarray}
In Appendix \ref{sec_c1}, we check that this relation is consistent
with the virial theorem (\ref{fp10}). The evolution of the moment of inertia is represented in Fig. \ref{i} for different values of the temperature. For $T>T_*$, the moment of inertia increases, for $T=T_*$ the moment of inertia is constant $I(t)=\mu r_{0}^2$ and for $T<T_*$ the moment of inertia decreases. This will become clear from the asymptotic behaviors.

For $t\rightarrow 0$, we get
\begin{eqnarray}
I(t)\simeq  \mu r_0^2+4D\mu (1-a)t,
\label{m4}
\end{eqnarray}
which can be rewritten
\begin{eqnarray}
\langle r^2\rangle \simeq   r_0^2+\frac{4k_B}{\xi\mu}(T-T_*)t.
\label{m5}
\end{eqnarray}
In that case, we have a normal diffusion with a gravity-modified
diffusion coefficient
\begin{eqnarray}
D(T)=\frac{k_B}{\xi\mu}(T-T_*).
\label{m5b}
\end{eqnarray}
For $T>T_*$ the variance increases with time while for $T<T_*$ it
decreases. This expression agrees with the naive virial theorem
(\ref{nv11}). Indeed, for small times, the probability for the
particle to reach $r=0$ is exponentially small so that the probability
is conserved and the naive virial theorem holds since there is no
Dirac peak.

For $t\rightarrow +\infty$, using the expansion (\ref{d5}),
we get
\begin{eqnarray}
I(t)\sim \frac{4\mu}{a\Gamma(a)}\left (\frac{r_0}{2}\right )^{2a}(Dt)^{1-a}.
\label{m7}
\end{eqnarray}
This corresponds to
\begin{eqnarray}
\langle r^2\rangle\sim \frac{4}{a\Gamma(a)}\left (\frac{r_0}{2}\right )^{2a}(Dt)^{1-a}.
\label{m8}
\end{eqnarray}
For $T>T_*$, i.e. $a<1$, the variance increases and goes to $+\infty$ for large
times. In that case, we have an anomalous diffusion $\langle
r^2\rangle \sim t^{\alpha}$ with an exponent $\alpha=1-T_*/T$. The
evolution is always sub-diffusive. The origin of the anomalous
diffusion is related to the fact that the particle can be trapped at
$r=0$ (and form a Dirac peak). For $T<T_*$, the variance decreases and
goes to $0$ for $t\rightarrow +\infty$.

\subsection{The most probable position}
\label{sec_mpp}

\begin{figure}
\begin{center}
\includegraphics[clip,scale=0.3]{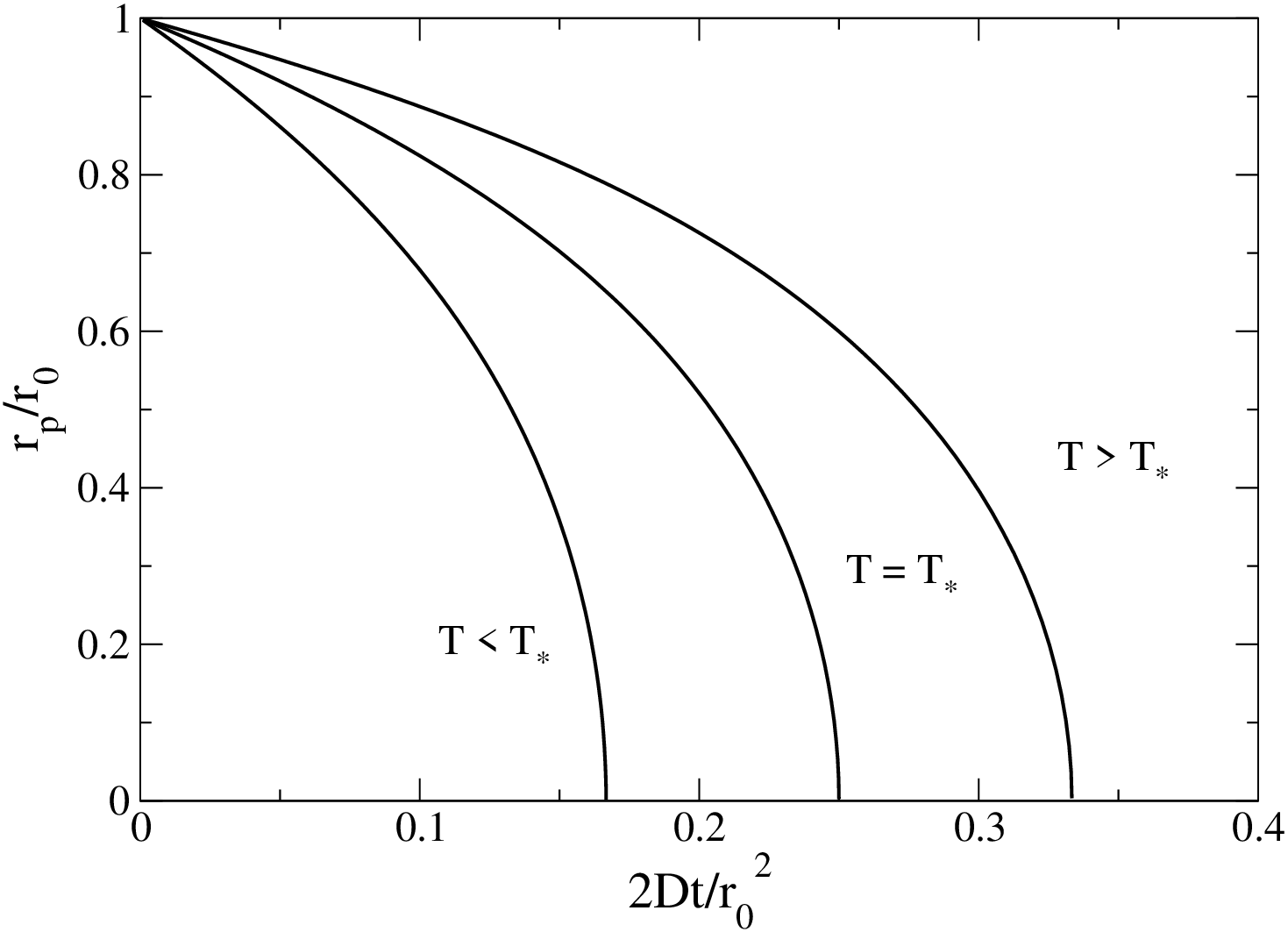}
\caption{Time evolution of $r_{P}$ for different values of the temperature (we have taken $T/T_*=1/2$, $T/T_*=1$ and $T/T_*=2$). }
\label{rp}
\end{center}
\end{figure}

The most probable value $r_{P}(t)$ of the distribution $P(r,t)$ is
obtained by maximizing $P(r,t)$, or equivalently $\ln P(r,t)$, with
respect to $r$. This gives
\begin{equation}
\frac{2aDt}{r_P r_0}+\frac{r_P}{r_0}=\frac{I'_a\left (\frac{r_Pr_0}{2Dt}\right )}{I_a\left (\frac{r_Pr_0}{2Dt}\right )}.
\label{pp1}
\end{equation}
Using the recurrence relation
\begin{equation}
I_{a}'(x)=I_{a+1}(x)+\frac{a}{x}I_{a}(x),
\label{pp2}
\end{equation}
we obtain
\begin{equation}
\frac{r_P}{r_0}=\frac{I_{a+1}\left (\frac{r_Pr_0}{2Dt}\right )}{I_a\left (\frac{r_Pr_0}{2Dt}\right )}.
\label{pp3}
\end{equation}
This equation can be rewritten in the parametric form
\begin{equation}
x=\frac{r_P r_0}{2Dt},\quad \frac{2Dt}{r_0^2}=\frac{I_{a+1}(x)}{xI_a(x)},
\label{pp4}
\end{equation}
or equivalently
\begin{equation}
\frac{r_P}{r_0}=\frac{I_{a+1}(x)}{I_a(x)},\quad \frac{2Dt}{r_0^2}=\frac{I_{a+1}(x)}{xI_a(x)},
\label{pp5}
\end{equation}
which gives $r_P(t)$. Using the asymptotic expansions of $I_{a}(x)$, we find that
\begin{equation}
\frac{r_{P}}{r_{0}}=1-\frac{2a+1}{2}\frac{2Dt}{r_{0}^2}, \quad (t\rightarrow 0),
\label{pp6}
\end{equation}
and
\begin{equation}
\frac{r_{P}}{r_{0}}=\left (\frac{a+2}{a+1}\right )^{1/2}\sqrt{1-t/t_c}, \quad (t\rightarrow t_c),
\label{pp7}
\end{equation}
where
\begin{equation}
\frac{2Dt_c}{r_0^2}=\frac{1}{2(a+1)}.
\label{pp8}
\end{equation}
Therefore, the radius $r_{P}(t)$ is decreasing for any temperature and
it goes to zero in a finite time $t_c(a)$ depending on the
temperature. Some curves are represented in Fig. \ref{rp}.

\begin{figure}
\begin{center}
\includegraphics[clip,scale=0.3]{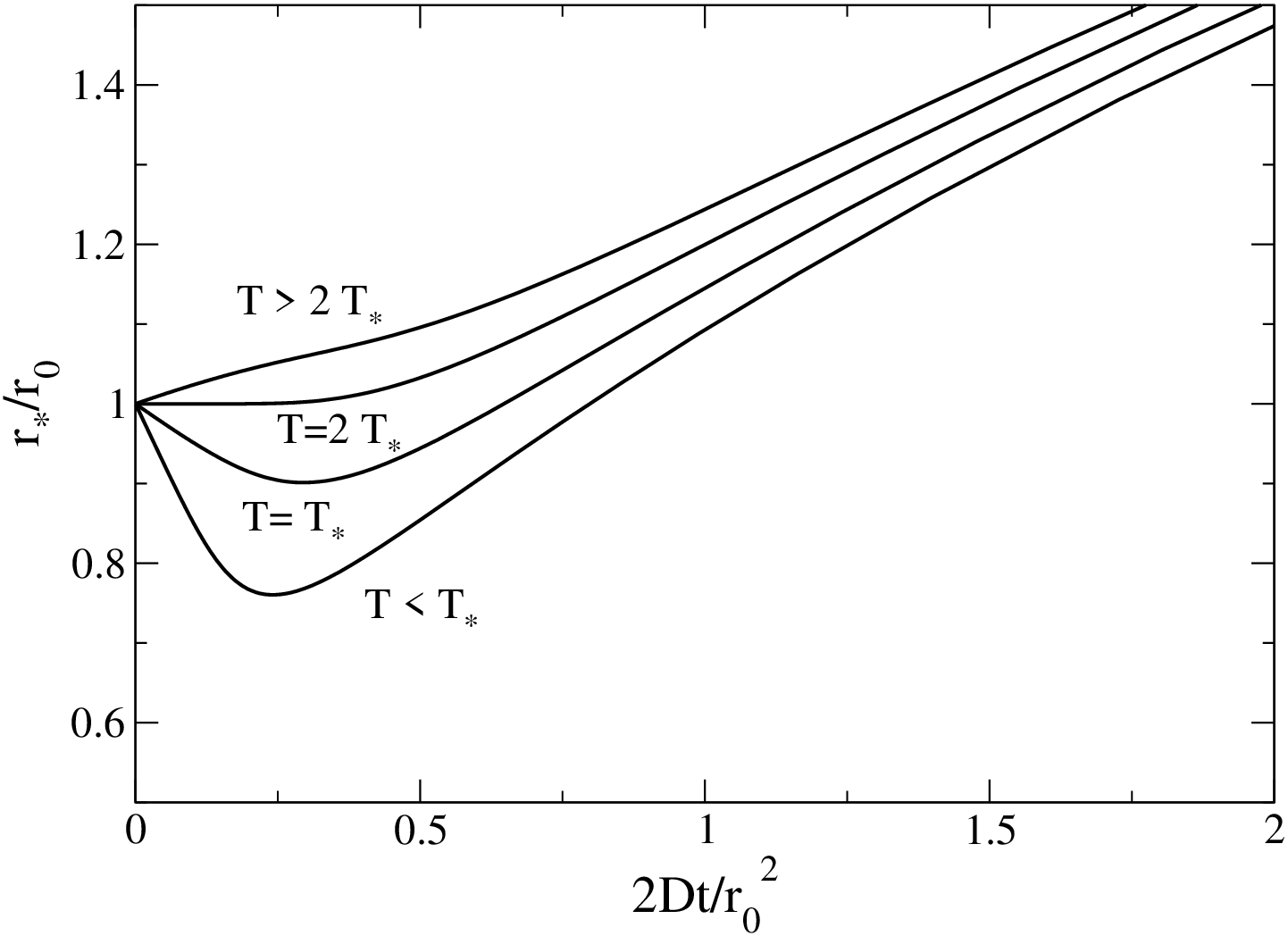}
\caption{Time evolution of $r_{*}$ for different values of the temperature (we have taken $T/T_*=1/2$, $T/T_*=1$, $T/T_*=2$ and  $T/T_*=4$). }
\label{rstar}
\end{center}
\end{figure}

The most probable value $r_{*}(t)$ of the radial distribution
$P(r,t)r$ is solution of
\begin{equation}
\frac{2(a-1)Dt}{r_* r_0}+\frac{r_*}{r_0}=\frac{I'_a\left (\frac{r_*r_0}{2Dt}\right )}{I_a\left (\frac{r_*r_0}{2Dt}\right )}.
\label{pp9}
\end{equation}
Using Eq. (\ref{pp2}), we get
\begin{equation}
\frac{r_*}{r_0}-\frac{2Dt}{r_{*}r_{0}}=\frac{I_{a+1}\left (\frac{r_*r_0}{2Dt}\right )}{I_a\left (\frac{r_*r_0}{2Dt}\right )}.
\label{pp10}
\end{equation}
This equation can be rewritten in the parametric form
\begin{equation}
x=\frac{r_* r_0}{2Dt},\quad \frac{2Dt}{r_0^2}=\frac{I_{a+1}(x)}{xI_a(x)}+\frac{1}{x^2},
\label{pp11}
\end{equation}
or equivalently
\begin{equation}
\frac{r_{*}}{r_{0}}=\frac{I_{a+1}(x)}{I_a(x)}+\frac{1}{x},\quad \frac{2Dt}{r_0^2}=\frac{I_{a+1}(x)}{xI_a(x)}+\frac{1}{x^2},
\label{pp12}
\end{equation}
which gives $r_{*}(t)$. Using the asymptotic expansions of $I_{a}(x)$, we find that
\begin{equation}
\frac{r_{*}}{r_{0}}=1-\frac{2a-1}{2}\frac{2Dt}{r_{0}^2}, \quad (t\rightarrow 0),
\label{pp13}
\end{equation}
and
\begin{equation}
r_{*}\sim\sqrt{2Dt},\qquad (t\rightarrow +\infty).
\label{pp14}
\end{equation}
Therefore, the radius $r_{*}(t)$ is always increasing for $T>2T_*$. For $T<2T_*$, it starts to decrease before finally increasing.   For $T=2T_*$, we can use the identities
\begin{equation}
I_{1/2}(x)=\sqrt{\frac{2}{\pi x}}\sinh(x),
\label{et1}
\end{equation}
\begin{equation}
I_{3/2}(x)=\sqrt{\frac{2}{\pi x}}\left \lbrack\cosh(x)-\frac{\sinh(x)}{x}\right \rbrack,
\label{et2}
\end{equation}
so that $r_*/r_0=1/\tanh(x)-1/x=L(x)$ (where $L(x)$ is the Langevin function) and $2Dt/r_0^2=1/(x\tanh(x))-1/x^2$. Using the asymptotic expansion of $\tanh(x)$ for $x\rightarrow 0$ and $x\rightarrow +\infty$, we find that
\begin{equation}
\frac{r_*}{r_0}\simeq 1+2e^{-\frac{r_0^2}{Dt}},\qquad (t\rightarrow 0).
\label{et3}
\end{equation}
Some curves are represented in Fig. \ref{rstar}.

\subsection{Reflecting boundary conditions}
\label{sec_rb}

In the previous sections, we have considered the case of absorbing boundary conditions at $r=0$. They lead to the formation of a Dirac peak with an amplitude $\chi_D(t)$ growing with time. We shall now consider the case of reflecting boundary conditions and determine their domain of existence. Reconsidering the calculations of Sec. \ref{sec_lp}, these boundary conditions correspond to solutions of the form
\begin{equation}
\phi_\lambda(r)= J_{-a}(\sqrt{\lambda/D}r),
\label{rb1}
\end{equation}
that diverge at the origin. Substituting Eq. (\ref{rb1}) in Eq. (\ref{lp3}) and repeating the calculations of Sec. \ref{sec_lp}, we obtain
\begin{equation}
P(r,t)=\left (\frac{r_0}{r}\right )^{a}\frac{1}{4\pi Dt}e^{-\frac{r_0^2+r^2}{4Dt}} I_{-a}\left (\frac{rr_0}{2Dt}\right ).
\label{rb2}
\end{equation}
According to identity (\ref{lp8}), this solution is valid only for $a<1$ (i.e. $T>T_*$). This is confirmed by considering the equivalent of $P(r,t)$ close to $r=0$:
\begin{equation}
P(r,t)\sim  \frac{1}{r^{\beta G m_1m_2}}\frac{4^a}{4\pi \Gamma(1-a)}\frac{1}{(Dt)^{1-a}}e^{-\frac{r_0^2}{4Dt}}.
\label{rb3}
\end{equation}
This distribution diverges at the origin and it is normalizable iff $T>T_*$. Since the distribution $P(r,t)$ diverges at the origin, we must replace Eq. (\ref{fp6}) by
\begin{equation}
\dot \chi(t)=-2\pi D \lim_{r\rightarrow 0} r\left (\frac{\partial P}{\partial r}+P\frac{\beta G m_1m_2}{r}\right ).   \label{rb4}
\end{equation}
Using Eq. (\ref{rb2}) and the expansion
\begin{equation}
I_{a}(x)=\frac{1}{\Gamma(a+1)}\left (\frac{x}{2}\right )^{a}+\frac{1}{\Gamma(a+2)}\left (\frac{x}{2}\right )^{a+2}+...,
\label{rb5}
\end{equation}
valid for $x\rightarrow 0$, we find that $\dot \chi=0$. Therefore, the normalization is conserved in time and there is no Dirac peak formation. 

These results are consistent with the van Kampen classification of singularities (see Appendix \ref{sec_vk}). For $T<T_*$ (i.e. $a>1$ or $\beta G m_1 m_2>2$), the singularity at $r=0$ behaves as an adhesive boundary. In that case, the solution is unique and no boundary condition has to be fixed by hand. It is given by Eq. (\ref{lp12}) leading to a Dirac peak  ($\dot\chi\neq 0$). On the other hand, for $T>T_*$ (i.e. $a<1$ or $\beta G m_1 m_2<2$), the singularity at $r=0$ behaves as a regular  boundary.  In that case, the boundary condition can be absorbing, reflecting or mixed. It has to be fixed by hand. In the previous sections, we considered  a purely absorbing boundary condition and, in the present section, we considered a purely reflecting boundary condition. In fact, for $a<1$ (i.e. $T>T_*$), the general solution can be written as a ``mixture'' of the two previous solutions
\begin{equation}
P(r,t)=\nu P_{+}(r,t)+(1-\nu) P_{-}(r,t),
\label{rb6}
\end{equation}
where $P_{+}(r,t)$ is the distribution (\ref{lp12}), $P_{-}(r,t)$ is the distribution (\ref{rb2}), and $\nu$ is a parameter taking values between $\nu=1$ (purely absorbing) and $\nu=0$ (purely reflecting).

\begin{figure}
\begin{center}
\includegraphics[clip,scale=0.3]{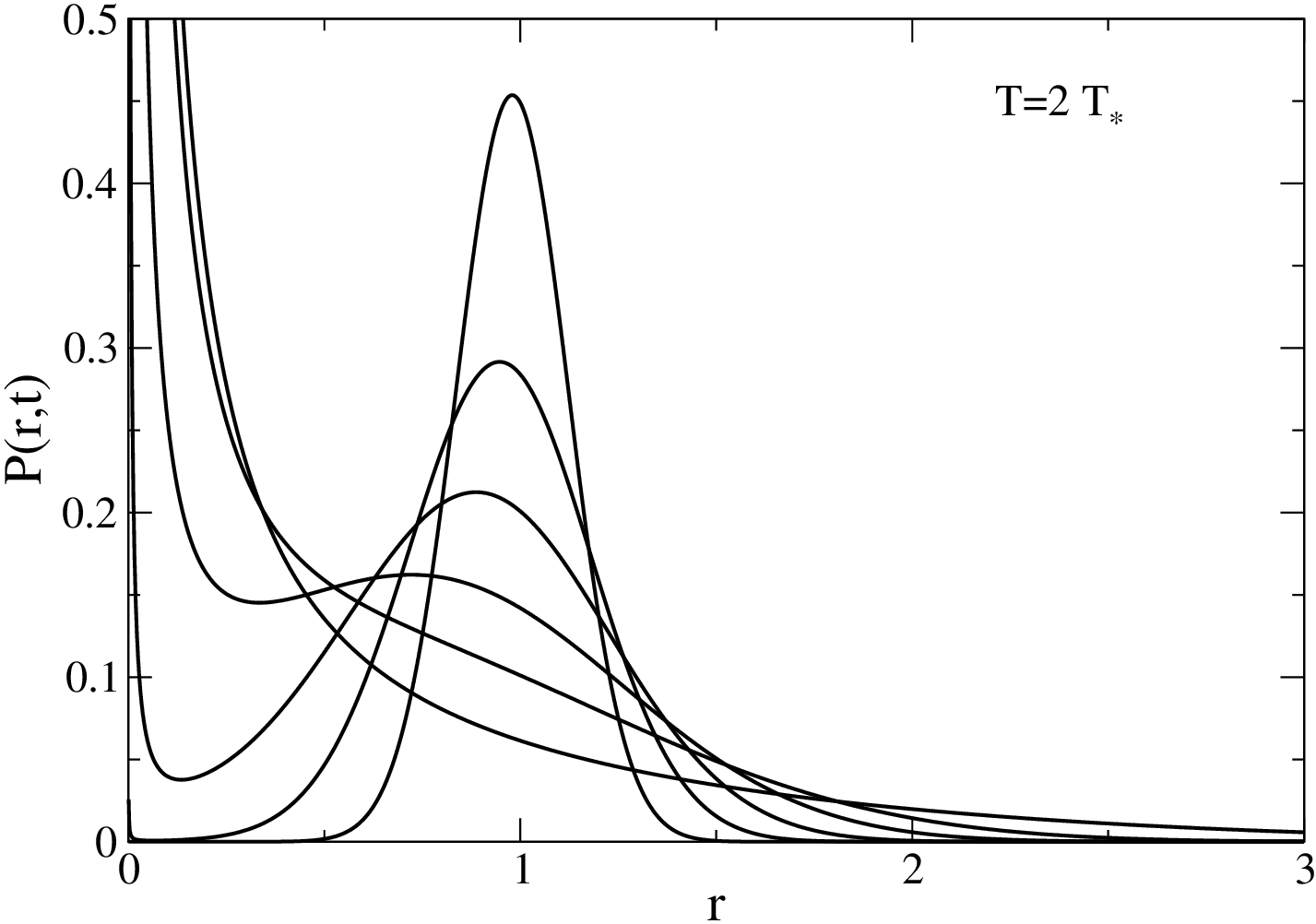}
\caption{Probability density  of finding the particles at a distance ${\bf r}$ from each other, at different times $t$ and for a temperature $T=2T_*$ ($a=1/2$). From top to bottom: $t=0.01, 0.025, 0.05, 0.1, 0.2, 1$.}
\label{pra0.5RB}
\end{center}
\end{figure}

Let us specifically consider the distribution (\ref{rb2}). It is
plotted in Fig. \ref{pra0.5RB} at different times for $T/T_*=2$
(corresponding to $a=1/2$).  Substituting Eq. (\ref{rb2}) in the expression 
(\ref{m1}) defining the moment of inertia of the reduced particle, and carrying
out the integrations, we find that
\begin{equation}
I(t)=\mu r_0^2+4D\mu(1-a)t.
\label{rb7}
\end{equation}
This expression agrees with the naive virial theorem
(\ref{nv10}). Indeed, in the present case $\chi(t)=1$ since there is
no Dirac peak formation, and the exact expression (\ref{fp10}) of the
virial theorem reduces to Eq. (\ref{nv10}).

\begin{figure}
\begin{center}
\includegraphics[clip,scale=0.3]{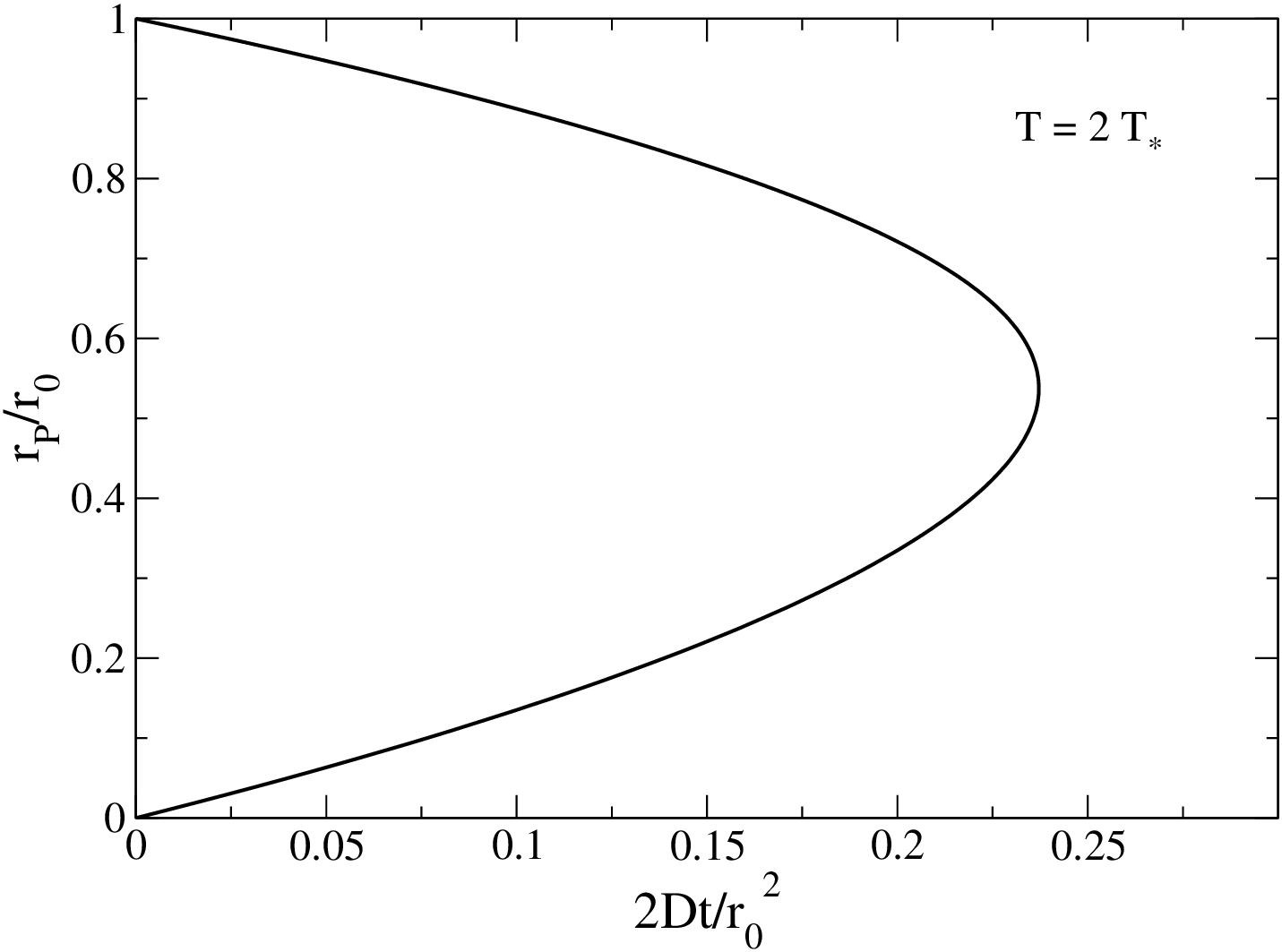}
\caption{Time evolution of $r_P$ for  $T=2T_*$ ($a=1/2$). The upper branch corresponds to the position of the local maximum and the lower branch to the position of the minimum.}
\label{rpta0.5RB}
\end{center}
\end{figure}

On the other hand, the most probable value $r_P(t)$ of the distribution $P(r,t)$ is obtained by maximizing $P(r,t)$, or equivalenly $\ln P(r,t)$ with respect to $r$. In fact, since this distribution diverges at $r=0$, the global maximum (infinite) is $r_P(t)=0$. However, for sufficiently short times, the distribution $P(r,t)$ also admits a local maximum $r_P^{(1)}(t)$ and a local minimum $r_P^{(2)}(t)$. Proceeding as in Sec. \ref{sec_mpp}, they are the solutions of the equation
\begin{equation}
\frac{4aDt}{r_P r_0}+\frac{r_P}{r_0}=\frac{I_{1-a}\left (\frac{r_Pr_0}{2Dt}\right )}{I_{-a}\left (\frac{r_Pr_0}{2Dt}\right )}.
\label{rb8}
\end{equation}
Setting $x=\frac{r_P r_0}{2Dt}$, this equation can be rewritten in the parametric form
\begin{equation}
\frac{r_P}{r_0}=\frac{I_{1-a}(x)}{I_{-a}(x)}-\frac{2a}{x},\quad \frac{2Dt}{r_0^2}=\frac{I_{1-a}(x)}{xI_{-a}(x)}-\frac{2a}{x^2},
\label{rb9}
\end{equation}
which gives $r_P(t)$. Using the asymptotic expansion of $I_{a}(x)$ for $x\rightarrow +\infty$, we find that
\begin{equation}
\frac{r_{P}^{(1)}}{r_{0}}=1-\frac{2a+1}{2}\frac{2Dt}{r_0^2}, \quad (t\rightarrow 0).
\label{rb10}
\end{equation}
The radius $r_{P}^{(1)}(t)$ of the local maximum is decreasing for any
temperature and it disappears at a time $t_{end}(a)$ which depends on
the temperature. The curve corresponding to $a=1/2$ is represented in
Fig. \ref{rpta0.5RB}.

\begin{figure}
\begin{center}
\includegraphics[clip,scale=0.3]{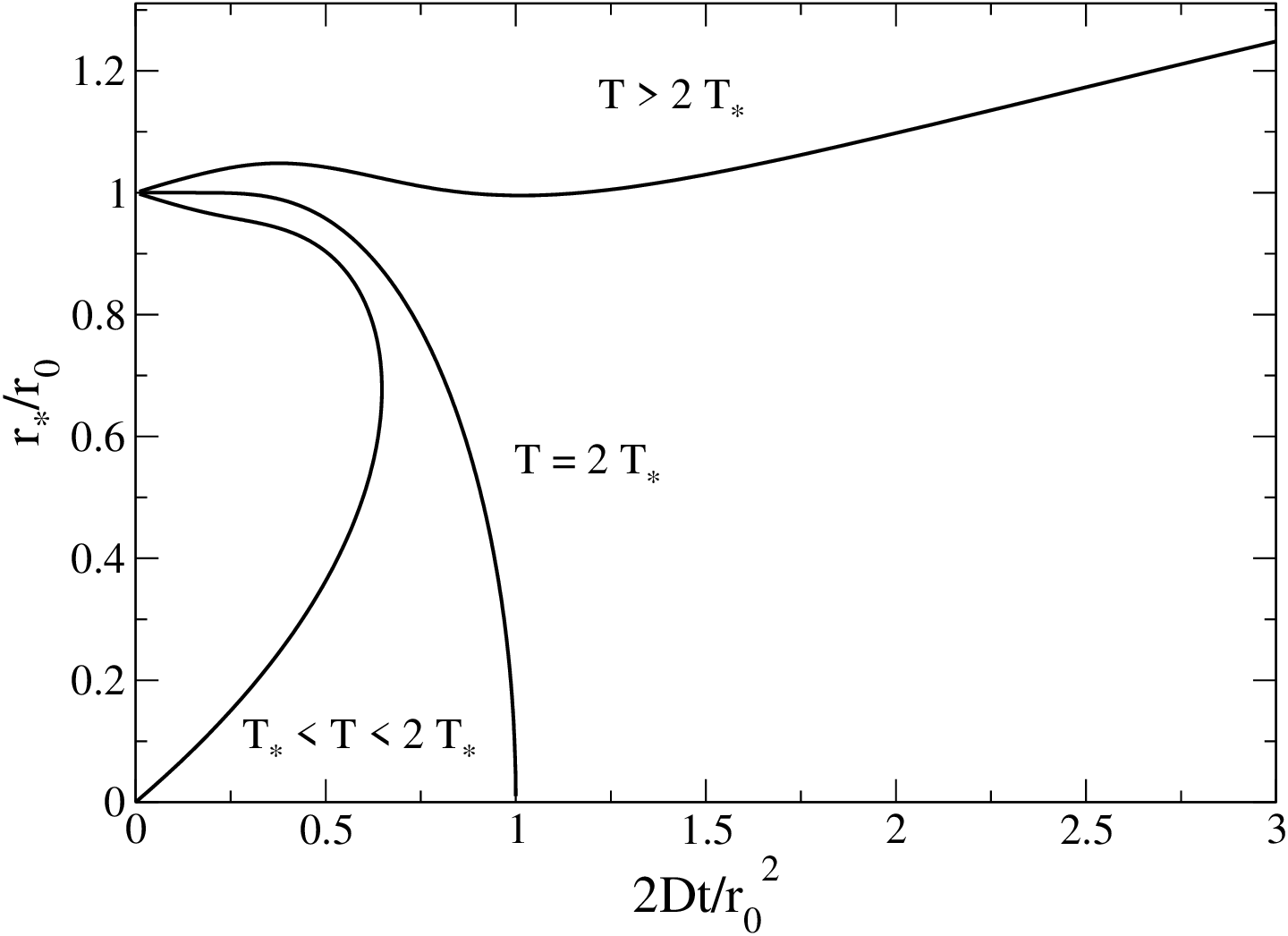}
\caption{Time evolution of $r_*$ for different values of $T$ (corresponding to $a=0.3$, $0.5$ and $0.7$). In the case $1/2< a<1$, the upper branch corresponds to the position of the local maximum and the lower branch to the position of the minimum.}
\label{rstarRB}
\end{center}
\end{figure}

Let us now consider the  most probable value $r_*(t)$ of the radial distribution $P(r,t)r$. This distribution diverges  at the origin $r=0$ for $1/2<a<1$,
vanishes at the origin for $0\le a<1/2$ and is finite at the origin for $a=1/2$. In the first case, the global maximum (infinite) is at $r_P=0$ but, for sufficiently short times, the distribution $P(r,t)r$ also admits a local maximum $r_P^{(1)}$ and a local minimum $r_P^{(2)}$. Proceeding as in Sec. \ref{sec_mpp}, they are the solutions of the equation
\begin{equation}
\frac{2(2a-1)Dt}{r_* r_0}+\frac{r_*}{r_0}=\frac{I_{1-a}\left (\frac{r_*r_0}{2Dt}\right )}{I_{-a}\left (\frac{r_*r_0}{2Dt}\right )}.
\label{rb11}
\end{equation}
Setting $x=\frac{r_* r_0}{2Dt}$, this equation can be rewritten in the parametric form
\begin{equation}
\frac{r_*}{r_0}=\frac{I_{1-a}(x)}{I_{-a}(x)}-\frac{2a-1}{x},\quad \frac{2Dt}{r_0^2}=\frac{I_{1-a}(x)}{xI_{-a}(x)}-\frac{2a-1}{x^2},
\label{rb12}
\end{equation}
which gives $r_*(t)$. Using the asymptotic expansion of $I_{a}(x)$ for $x\rightarrow +\infty$, we find that
\begin{equation}
\frac{r_{*}^{(1)}}{r_{0}}=1+\frac{1-2a}{2}\frac{2Dt}{r_0^2}, \quad (t\rightarrow 0).
\label{rb14}
\end{equation}
For $1/2<a<1$, the radius $r_{*}^{(1)}(t)$ of the local maximum is decreasing and it disappears at a time $t_{end}(a)$ which depends on the temperature. For $0\le a<1/2$, the radius $r_{*}^{(1)}(t)$ increases initially and behaves for large times as
\begin{equation}
r_{*}\sim \sqrt{2(1-2a)Dt}, \quad (t\rightarrow +\infty).
\label{rb13}
\end{equation}
For $a=1/2$, we can use the identities (\ref{et1}) and
\begin{equation}
I_{-1/2}(x)=\sqrt{\frac{2}{\pi x}}\cosh(x),
\label{idbes1}
\end{equation}
so that $r_*/r_0=\tanh(x)$ and $2Dt/r_0^2=\tanh(x)/x$. Using the asymptotic expansions of $\tanh(x)$ for $x\rightarrow 0$ and $x\rightarrow +\infty$, we find that
\begin{equation}
\frac{r_*}{r_0}\simeq 1-2e^{-\frac{r_0^2}{Dt}},\qquad (t\rightarrow 0),
\label{idbes2}
\end{equation}
\begin{equation}
\frac{r_*}{r_0}\simeq \sqrt{3}\left (1-\frac{2Dt}{r_0^2}\right )^{1/2},\qquad (t\rightarrow \frac{r_0^2}{2D}),
\label{idbes3}
\end{equation}
establishing $t_{end}=t_c=\frac{r_0^2}{2D}$ for $a=1/2$. Some curves are represented in Fig. \ref{rstarRB}. 

\section{The case of a bounded domain}
\label{sec_b}

In the previous sections, we considered the case where the particles
are free to move in an infinite domain. We shall now consider the
case where the particles are confined in a bounded domain. As an
idealization, we shall consider that the reduced particle evolves in a
spherical box of radius $R$. We first look for the existence of an
equilibrium state. The steady solution of the Fokker-Planck equation
(\ref{fp1}) satisfies
\begin{equation}
\nabla P_{eq}+\beta G m_1 m_2 P_{eq}\frac{\bf r}{r^2}={\bf 0},
\label{b1}
\end{equation}
which is integrated into
\begin{equation}
P_{eq}=\frac{A}{r^{\beta Gm_1m_2}}.
\label{b2}
\end{equation}
It is clear that this distribution is normalizable iff $\beta
Gm_1m_2<2$ so that an equilibrium state exists iff
$T>T_*$. For $T<T_*$, the singularity at $r=0$ is
adhesive and a Dirac peak grows, ultimately absorbing the particle
with probability one. For $T>T_*$, the singularity at $r=0$ is regular
and its nature (absorbing, reflecting or mixed) has to be fixed by
hand. In the case of an absorbing boundary, we ultimately obtain a
Dirac peak of amplitude $\chi_D(+\infty)=1$, so that there is no
equilibrium state outside the Dirac. In the case of a reflecting
boundary, there is no Dirac peak and the system reaches an equilibrium
state, given by Eq.  (\ref{b2}), with normalization $\int P_{eq}\,
d{\bf r}=1$. In the mixed case, we ultimately obtain a Dirac peak with
amplitude $\chi_D(+\infty)=\nu$ and an equilibrium state outside the
Dirac, given by Eq.  (\ref{b2}), with normalization $\int P_{eq}\,
d{\bf r}=1-\nu$.

Let us now solve the Fokker-Planck equation (\ref{fp1}) in a bounded domain.
In that case, we need to impose the boundary condition
\begin{equation}
\frac{\partial P}{\partial r}+\beta P\frac{\partial U}{\partial r}=0,
\label{b3}
\end{equation}
in $r=R$ meaning that there is no flux of probability at the boundary. In terms of the eigenfunction $\phi$ defined in Eqs. (\ref{s3}) and (\ref{s8}), this can be rewritten
\begin{equation}
\frac{\phi'(R)}{\phi(R)}=-\frac{1}{2}\beta U'(R).
\label{b4}
\end{equation}
For the potential $U=Gm_1m_2\ln r$, we obtain
\begin{equation}
\frac{\phi'(R)}{\phi(R)}=-\frac{\beta Gm_1m_2}{2R}.
\label{b5}
\end{equation}
This boundary condition implies that the eigenvalues are quantized.  

Let us first consider purely absorbing boundary conditions at $r=0$. Making the change of notations $\lambda\rightarrow D\lambda^2$, the eigenfunctions that are solution of Eq. (\ref{lp4}) and that are finite at the origin are given by
\begin{equation}
\phi_n(r)=A_n J_a(\lambda_n r).
\label{b6}
\end{equation}
Substituting this solution in Eq. (\ref{b5}), we find that the eigenvalues are determined by
\begin{equation}
\frac{\lambda_n R J_{a}'(\lambda_n R)}{J_{a}(\lambda_n R)}=-a.
\label{b7}
\end{equation}
Using the recurrence relation
\begin{equation}
J'_a(x)=J_{a-1}(x)-\frac{a}{x}J_{a}(x),
\label{b7b}
\end{equation}
the foregoing equation can be rewritten
\begin{equation}
J_{a-1}(\lambda_n R)=0,
\label{b7c}
\end{equation}
with $\lambda_n\neq 0$. Therefore, the eigenvalues $R\lambda_{n}(a)$ are the zeros of $J_{a-1}(x)$. Using the general identity
\begin{eqnarray}
\int_{0}^{R} J_a^2(\lambda r)r\, dr=\frac{R^2}{2}\left\lbrack J_a'(\lambda R)^2+\left (1-\frac{a^2}{\lambda^2R^2}\right )J_a^2(\lambda R)\right\rbrack,\nonumber\\
\label{b9}
\end{eqnarray}
together with the relation  (\ref{b7}), we obtain
\begin{eqnarray}
\int_{0}^{R} J_a^2(\lambda_n r)r\, dr=\frac{R^2}{2}J_a^2(\lambda_n R).
\label{b10}
\end{eqnarray}
Therefore, the normalized eigenfunctions are
\begin{equation}
\phi_n(r)=\frac{1}{\sqrt{\pi}RJ_a(\lambda_n R)} J_a(\lambda_n r).
\label{b11}
\end{equation}
Finally, using Eq. (\ref{gs4}), the solution of the Fokker-Planck equation (\ref{fp1}) in a bounded domain can be written
\begin{equation}
P(r,t)=\left (\frac{r_0}{r}\right )^{a}\sum_n e^{-D\lambda_n^2 t}\phi_{n}(r_{0})\phi_{n}(r).
\label{b8}
\end{equation}
We can now redo the preceding analysis except that the results will
be less explicit since they will be expressed in the form of series.

The probability $\chi_D(t)$ for the particle to have formed a Dirac peak at time $t$ is
given by Eq. (\ref{fp8}). Using Eq. (\ref{b8}) and the equivalent
\begin{equation}
J_{n}(x)\sim \frac{1}{\Gamma(n+1)}\left (\frac{x}{2}\right )^n, \qquad (x\rightarrow 0),
\label{mar1}
\end{equation}
we obtain
\begin{eqnarray}
P(0,t)=\frac{1}{\pi R^2}\left (\frac{r_0}{2}\right )^a \frac{1}{\Gamma(a+1)}
\sum_n \lambda_n^a\frac{J_a(\lambda_n r_0)}{J_a^2(\lambda_n R)}e^{-D\lambda_n^2 t}.\nonumber\\
\label{mar2}
\end{eqnarray}
Therefore, the probability for the particle to have formed a Dirac peak at time $t$ can be written
\begin{eqnarray}
\chi_D(t)=1-\sum_n B_n e^{-D\lambda_n^2 t},
\label{mar3}
\end{eqnarray}
with
\begin{eqnarray}
B_n=\frac{2\beta G m_1m_2}{R^2}\left (\frac{r_0}{2}\right )^a \frac{1}{\Gamma(a+1)}
 \lambda_n^{a-2}\frac{J_a(\lambda_n r_0)}{J_a^2(\lambda_n R)}.\nonumber\\
\label{mar4}
\end{eqnarray}
For $t\rightarrow +\infty$, we obtain
\begin{eqnarray}
\chi_D(t)\simeq 1- B_1 e^{-D\lambda_1^2 t},
\label{mar5}
\end{eqnarray}
so that the probability that the particle has not formed a Dirac at
time $t$ decreases {\it exponentially} as $e^{-D\lambda_1^2 t}$
instead of algebraically in an unbounded domain (see
Sec. \ref{sec_d}). The exponential decay is controlled by the first
eigenvalue $\lambda_1(T)$ (fundamental) of the Schr\"odinger equation.
As we have seen, $\lambda_1(a)R$ is the first zero of
$J_{a-1}(x)$. This is valid for any
temperature.  It is instructive,
however, to determine the asymptotic behavior of $\lambda_1$ for
$a\rightarrow 0$ (i.e. $T\rightarrow +\infty$). In that limit, $\lambda_1\rightarrow 0$. Substituting the expansion
\begin{eqnarray}
J_{n}(x)=\left (\frac{x}{2}\right )^n\left\lbrack \frac{1}{\Gamma(n+1)}-\frac{1}{\Gamma(n+2)}\left (\frac{x}{2}\right )^2+...\right\rbrack,\nonumber\\
\label{mar6}
\end{eqnarray}
valid for $x\rightarrow 0$ in Eq. (\ref{b7c}), we obtain
\begin{eqnarray}
(\lambda_1 R)^2\sim 4a\sim \frac{2Gm_1m_2}{k_B T}.
\label{mar7}
\end{eqnarray}
On the other hand, in this limit,
$B_1\rightarrow 1$. Therefore, Eq. (\ref{mar5}) becomes
\begin{eqnarray}
\chi_D(t)\simeq 1- e^{-\frac{2DGm_1m_2}{k_B TR^2} t}.
\label{mar8}
\end{eqnarray}
Using Eq. (\ref{fp2}), it can be rewritten
\begin{eqnarray}
\chi_D(t)\simeq 1- e^{-\frac{2GM}{\xi R^2} t}.
\label{mar9}
\end{eqnarray}
For $T=T_*$ (i.e. $a=1$), we
find that $\lambda_1 R= j_{0,1}$ where $j_{0,1}\simeq
2.40482...$ is the first zero of $J_{0}(x)$. Finally, for
$T\rightarrow 0$, we find that 
\begin{eqnarray}
\lambda_1 R\sim a\sim \frac{Gm_1m_2}{2k_B T}.
\label{marneq}
\end{eqnarray}

Let us now consider purely reflecting boundary conditions at $r=0$. Making the change of notations $\lambda\rightarrow D\lambda^2$, the eigenfunctions that are solution of Eq. (\ref{lp4}) and that diverge at the origin are given by
\begin{equation}
\phi_n(r)=A_n J_{-a}(\lambda_n r).
\label{b6bis}
\end{equation}
Substituting this solution in Eq. (\ref{b5}), we find that the eigenvalues are determined by
\begin{equation}
\frac{\lambda_n R J_{-a}'(\lambda_n R)}{J_{-a}(\lambda_n R)}=-a.
\label{b7bis}
\end{equation}
Using the recurrence relation
\begin{equation}
J'_a(x)=-J_{a+1}(x)+\frac{a}{x}J_{a}(x),
\label{b7bbis}
\end{equation}
the foregoing equation can be rewritten
\begin{equation}
J_{1-a}(\lambda_n R)=0,
\label{b7cbis}
\end{equation}
with $\lambda_n\neq 0$. Therefore, the eigenvalues $R\lambda_{n}(a)$
are the zeros of $J_{1-a}(x)$. Using the general identity (\ref{b9})
together with the relation (\ref{b7bis}), we find that the normalized
eigenfunctions are
\begin{equation}
\phi_n(r)=\frac{1}{\sqrt{\pi}RJ_{-a}(\lambda_n R)} J_{-a}(\lambda_n r).
\label{b11bis}
\end{equation}
This expression is valid for $\lambda_n\neq 0$. We must also add the eigenmode corresponding to $\lambda_0=0$ whose normalized expression is
\begin{equation}
\phi_0(r)=\sqrt{\frac{1-a}{\pi}}\frac{1}{R}\left (\frac{R}{r}\right )^a.
\label{b11bist}
\end{equation}
Finally, using Eq. (\ref{gs4}), the solution of the Fokker-Planck equation (\ref{fp1}) in a bounded domain can be written
\begin{equation}
P(r,t)=P_{eq}(r)+\left (\frac{r_0}{r}\right )^{a}\sum_n e^{-D\lambda_n^2 t}\phi_{n}(r_{0})\phi_{n}(r),
\label{b8bis}
\end{equation}
where
\begin{equation}
P_{eq}(r)=\frac{1-a}{\pi R^2}\left (\frac{R}{r}\right )^{2a},
\label{b8bistr}
\end{equation}
is the equilibrium distribution and the series run over $n\ge 1$. 
Using Eq. (\ref{b8bis}) and the equivalent (\ref{mar1}), we obtain for
$r\rightarrow 0$:
\begin{eqnarray}
P(r,t)- P_{eq}(r)\sim\frac{1}{\pi R^2}\frac{1}{r^{2a}} \frac{(2r_0)^a}{\Gamma(1-a)}\nonumber\\
\times\sum_n \frac{1}{\lambda_n^a}\frac{J_{-a}(\lambda_n r_0)}{J_{-a}^2(\lambda_n R)}e^{-D\lambda_n^2 t}.
\label{mar2bis}
\end{eqnarray}
The distribution diverges at the origin and it is normalizable iff
$a<1$, i.e. $T>T_*$. On the other hand, using Eqs. (\ref{b8bis}) and
(\ref{rb4}), we find that $\dot\chi=0$ so that there is no Dirac peak in that case. For $t\rightarrow +\infty$, we get
\begin{equation}
P(r,t)-P_{eq}(r)\sim \left (\frac{r_0}{r}\right )^{a} \phi_{1}(r_{0})\phi_{1}(r)e^{-D\lambda_1^2 t},
\label{b8bisinf}
\end{equation}
so that the distribution converges exponentially rapidly towards the equilibrium state as $e^{-D\lambda_1^2 t}$. The exponential  relaxation time is controlled by the first eigenvalue $\lambda_1(a)$ (fundamental) of the Schr\"odinger equation. As we have seen, $R\lambda_1(a)$ is the first zero of $J_{1-a}(x)$. This is valid for any temperature $T>T_*$. For $T\rightarrow +\infty$ (i.e. $a\rightarrow 0$), we find that $\lambda_1 R\rightarrow j_{1,1}$ where $j_{1,1}\simeq 3.83171...$ is the first zero of $J_{1}(x)$. For $T= T_*$ (i.e. $a= 1$), we find that $\lambda_1 R= j_{0,1}$ where $j_{0,1}\simeq 2.40482...$ is the first zero of $J_{0}(x)$.

{\it Remark:} we could also consider the case of $N=2$
self-gravitating Brownian particles with a short-range
regularization. This could be due to a softened potential
$U=Gm_1m_2\ln(\sqrt{r^2+\epsilon^2})$, to a hard core $a$ or to an
exclusion principle such as the Pauli exclusion principle for fermions
in quantum mechanics. In that case, there is no Dirac peak
formation. There exists an equilibrium state for any temperature in a
box and for $T<T_*$ in an infinite domain \cite{epjb}. For low
temperatures, the two particles are at a typical distance $a$ from
each other so that the equilibrium state is controlled by the
small-scale cut-off. For high temperatures, the particles are at a
typical distance $R$ (the box radius) from each other in a bounded
domain or tend to ``evaporate'' to infinity in an unbounded domain.

\section{By-product: post-collapse of the Smoluchowski-Poisson system}
\label{sec_bp}

There is an interesting by-product of the previous study. Indeed, the
preceding analysis can be used to obtain new results concerning the
post-collapse dynamics of the Smoluchowski-Poisson system (or
Keller-Segel model). In $d=2$ dimensions and for
$T<T_c=GMm/(4k_B)$, it is known that the Smoluchowski-Poisson system
forms a Dirac peak of mass $M_0=(T/T_c)M$ in a finite time $t_{coll}$
\cite{sc}. For $t>t_{coll}$ the Dirac continues to grow by accretion
of the surrounding matter\footnote{This post-collapse regime has been
studied in \cite{post} for $d>2$. For $d=2$, it is more difficult to
study except in the large time limit $t\rightarrow +\infty$ that we
consider here. Note that the virial theorem given in
\cite{virial} is not valid anymore when a Dirac peak is
formed (in the post-collapse regime). The reason is the same as the
one given in Sec. \ref{sec_fp}. The correct form of
the virial theorem in that case is given in Appendix \ref{sec_post}.}. For
$T=T_c^-$ or for $t\rightarrow +\infty$ and any $T<T_c$, the Dirac
peak has accreted most of the mass. As a result, the system is formed
by a Dirac peak of mass $M_0(t)\simeq M$ surrounded by a dilute halo
containing the remaining mass $\epsilon=M-M_0(t)$. Now, it is possible to
neglect the self-gravity of the halo and consider that the particles
of the halo are only subject to the gravity of the central peak of
mass $\sim M$. Therefore, the evolution of the halo density is
governed by a Fokker-Planck equation of the form
\begin{equation}
\xi{\partial \rho\over \partial t}=\nabla\cdot \left (\frac{k_B T}{m}\nabla \rho+\rho\frac{GM {\bf r}}{r^2}\right ). \label{bp1}
\end{equation}
This equation has been studied in \cite{post} in a box. However, it
was not realized that the corresponding eigenvalue equation could be
solved analytically in $d=2$. Indeed, Eq. (\ref{bp1}) is equivalent to
Eq. (\ref{fp1}) up to a change of notations and we can therefore apply
the results of the previous sections. It suffices to define
$D=k_BT/(m\xi)$ and $U=GMm\ln r$. Then, we can use the results of the previous sections with now
\begin{equation}
a=\frac{\beta GMm}{2}=\frac{2T_c}{T}.\label{bp2b}
\end{equation}

In an infinite domain, using Eq. (\ref{d6}), we find that the mass of the Dirac peak saturates to
$M$ algebraically rapidly as
\begin{equation}
1-\frac{M_0(t)}{M}\sim t^{-a}. \label{bp2}
\end{equation}
In a bounded domain, using Eq. (\ref{mar5}),  we find that the mass
of the Dirac peak  saturates to $M$ exponentially rapidly as
\begin{equation}
1-\frac{M_0(t)}{M}\sim e^{-D\lambda_1^2(a)t}, \label{bp3}
\end{equation}
where $R\lambda_1(a)$ is the first zero of $J_{a-1}(x)$. This result was
previously found in \cite{post} but the exponential rate (eigenvalue)
was not obtained explicitly (except in the asymptotic limit
$T\rightarrow 0$). Note finally that for $T=0$, $M_0(t)$ saturates to $M$ in a finite time
\begin{equation}
t_{end}=\frac{R^2}{2GM}, \label{bp4}
\end{equation}
corresponding to the deterministic collapse of the outer mass annulus initially at $r=R$ (see Eq. (\ref{z4}) with $k=GM/\xi$).

\section{The logarithmic Fokker-Planck equation in $d$ dimensions}
\label{sec_lfp}

In this section, we briefly generalize the previous results to the logarithmic Fokker-Planck equation in $d$ dimensions
\begin{equation}
\xi{\partial P\over \partial t}=\nabla\cdot \left (\frac{k_B T}{\mu}\nabla P+P\frac{Gm_1m_2}{\mu}\frac{{\bf r}}{r^2}\right ). \label{lfp1}
\end{equation}
The previous results are recovered for $d=2$. Taking the time derivative of the moment of inertia of the reduced particle (\ref{fp9}), and using Eq. (\ref{lfp1}), we obtain after an integration by parts
\begin{equation}
\xi\dot I=-2\int {\bf r}\cdot \left (k_B T\nabla P+P Gm_1m_2\frac{\bf r}{r^2}\right )\, d{\bf r}.
 \label{lfp2}
 \end{equation}
It can be shown, using the following expressions for $P({\bf r},t)$, that the boundary terms at $r=0$ and $r=+\infty$ vanish. Integrating again by parts and introducing the notation $\chi(t)=\int P({\bf r},t)\, d{\bf r}$ taking into account the possibility that the normalization of $P({\bf r},t)$ is not conserved (due to the formation of a Dirac peak at ${\bf r}={\bf 0}$), we obtain
\begin{equation}
\xi\dot I=2k_B T d \chi(t)-2G m_1m_2\chi(t).
 \label{lfp3}
 \end{equation}
Finally, this can be rewritten
\begin{equation}
\frac{1}{2d}\xi\frac{dI}{dt}=  \chi(t)k_B (T-T_*),
 \label{lfp4}\end{equation}
where we have introduced the critical temperature
\begin{equation}
k_BT_*=\frac{Gm_1m_2}{d}.
 \label{lfp5}
 \end{equation}
We emphasize that this relation is valid whether $P({\bf r},t)$ is spherically symmetric or not.

We now consider a spherically symmetric evolution. Using the notations of Eq. (\ref{fp2}), we can write the logarithmic Fokker-Planck equation (\ref{lfp1}) in the form
\begin{eqnarray}
\frac{\partial P}{\partial t}=\frac{1}{r^{d-1}}\frac{\partial}{\partial r}\left\lbrace Dr^{d-1}\left (\frac{\partial P}{\partial r}+P\frac{\beta G m_1 m_2}{r}\right )\right\rbrace.
\label{lfp6}
\end{eqnarray}
The time derivative of the normalization is
\begin{equation}
\dot \chi(t)=-S_d D \lim_{r\rightarrow 0} r^{d-1}\left (\frac{\partial P}{\partial r}+P\frac{\beta G m_1m_2}{r}\right ).   \label{lfp7}
\end{equation}
The Fokker-Planck equation (\ref{lfp6}) can be transformed into a Schr\"odinger equation (\ref{s6}) with a potential
\begin{eqnarray}
V(r)=-\frac{D}{r^2}\left \lbrack a^2-\frac{(d-2)^2}{4}\right \rbrack,
\label{lfp8}
\end{eqnarray}
where we have defined
\begin{eqnarray}
a=\frac{\beta G m_1 m_2-(d-2)}{2}.
\label{lfp9}
\end{eqnarray}
When $d>2$, we see that $a=0$ at the new critical temperature
\begin{eqnarray}
k_B T'_*=\frac{G m_1 m_2}{d-2}.
\label{lfp10}
\end{eqnarray}
In the following discussion, it is implicit that $T_*'=+\infty$ if
$d\le 2$. The eigenvalue equation (\ref{s9}) takes the form
\begin{eqnarray}
r^2\phi''+(d-1)r\phi'+\left\lbrack\frac{\lambda}{D}r^2-a^2+\frac{(d-2)^2}{4}\right\rbrack\phi=0,\nonumber\\
\label{lfp11}
\end{eqnarray}
and it can be solved analytically in terms of Bessel functions.

Let us first consider the solution
\begin{eqnarray}
\phi_+(r)=r^{\frac{2-d}{2}}J_{|a|}(\sqrt{\lambda/D}r).
\label{lfp12}
\end{eqnarray}
Repeating the calculations of Sec. \ref{sec_lp}, we find that the solution of the logarithmic Fokker-Planck equation (\ref{lfp6}) with the initial condition (\ref{gs1}) is
\begin{equation}
P_+(r,t)=\frac{1}{r^{d-2}}\left (\frac{r_0}{r}\right )^{a}\frac{1}{2S_d Dt}e^{-\frac{r_0^2+r^2}{4Dt}} I_{|a|}\left (\frac{rr_0}{2Dt}\right ).
\label{lfp13}
\end{equation}
We need to distinguish two cases. For $a>0$ (i.e. $\beta G m_1 m_2>d-2$ or $T<T'_*$), the behavior of the distribution close to $r=0$ is
\begin{equation}
P_+(r,t)\sim  \frac{1}{r^{d-2}}\left (\frac{r_0}{2}\right )^{2a}\frac{1}{2S_d \Gamma(a+1)}\frac{1}{(Dt)^{a+1}}e^{-\frac{r_0^2}{4Dt}}.
\label{lfp14}
\end{equation}
Substituting this equivalent in Eq. (\ref{lfp7}), we find that
\begin{equation}
\dot\chi_+(t)=-D\left (\frac{r_0}{2}\right )^{2a}\frac{1}{\Gamma(a)}\frac{1}{(Dt)^{a+1}}e^{-\frac{r_0^2}{4Dt}}.
\label{lfp15}
\end{equation}
In that case, we have an absorbing boundary condition at $r=0$ and the growth of a Dirac peak. For $a<0$ (i.e. $\beta G m_1 m_2<d-2$ or $T>T'_*$), the behavior of the distribution close to $r=0$ is
\begin{equation}
P_+(r,t)\sim  \frac{1}{r^{\beta G m_1 m_2}}\frac{1}{4^{|a|}}\frac{1}{2S_d \Gamma(1+|a|)}\frac{1}{(Dt)^{|a|+1}}e^{-\frac{r_0^2}{4Dt}}.
\label{lfp16}
\end{equation}
Substituting this equivalent in Eq. (\ref{lfp7}), we find that $\dot\chi_+=0$. In that case, the normalization condition is conserved and there is no Dirac peak.

Let us now consider the solution
\begin{eqnarray}
\phi_-(r)=r^{\frac{2-d}{2}}J_{-|a|}(\sqrt{\lambda/D}r).
\label{lfp17}
\end{eqnarray}
Repeating the calculations of Sec. \ref{sec_lp}, we find that the solution of the logarithmic Fokker-Planck equation (\ref{lfp6}) with initial condition (\ref{gs1}) is
\begin{equation}
P_-(r,t)=\frac{1}{r^{d-2}}\left (\frac{r_0}{r}\right )^{a}\frac{1}{2S_d Dt}e^{-\frac{r_0^2+r^2}{4Dt}} I_{-|a|}\left (\frac{rr_0}{2Dt}\right ),
\label{lfp18}
\end{equation}
provided that $|a|<1$, according to identity (\ref{lp8}). We need to
distinguish two cases. For $0<a<1$ (i.e. $d-2<\beta G m_1 m_2<d$ or
$T_*<T<T'_*$), the behavior of the distribution close to $r=0$ is
\begin{equation}
P_-(r,t)\sim  \frac{1}{r^{\beta G m_1 m_2}}\frac{4^{a}}{2S_d \Gamma(1-a)}\frac{1}{(Dt)^{1-a}}e^{-\frac{r_0^2}{4Dt}}.
\label{lfp19}
\end{equation}
Substituting this equivalent in Eq. (\ref{lfp7}), we find that $\dot\chi_-=0$. In that case, the normalization condition is conserved and there is no Dirac peak. For $-1<a<0$ (i.e. $d-4<\beta G m_1 m_2<d-2$ or $T'_*<T<T''_*$ with $k_B T''_*=Gm_1 m_2/(d-4)$), the behavior of the distribution close to $r=0$ is
\begin{equation}
P_-(r,t)\sim  \frac{1}{r^{d-2}}\left (\frac{2}{r_0}\right )^{2|a|}\frac{1}{2S_d \Gamma(1-|a|)}\frac{1}{(Dt)^{1-|a|}}e^{-\frac{r_0^2}{4Dt}}.
\label{lfp20}
\end{equation}
Substituting this equivalent in Eq. (\ref{lfp7}), we find that $\dot\chi_->0$ which is not physically possible (the normalization of $P(r,t)$ can decrease if the particle is absorbed at $r=0$, but it cannot spontaneously increase). Therefore, this solution must be rejected.

These results are consistent with the van Kampen classification of singularities (see Appendix \ref{sec_vk}). For $T<T_*$ (i.e. $a>1$ or $\beta G m_1 m_2>d$), the singularity at $r=0$ behaves as an adhesive boundary. In that case, the solution is unique and no boundary condition has to be fixed by hand. It is given by Eq. (\ref{lfp13}) leading to a Dirac peak  ($\dot\chi_+\neq 0$). For $T_*<T<T'_*$ (i.e. $0<a<1$ or $d-2<\beta G m_1 m_2<d$), the singularity at $r=0$ behaves as a regular  boundary.  In that case, the boundary condition can be absorbing, reflecting or mixed. It has to be fixed by hand. The general solution can be written as a ``mixture'' (\ref{rb6}) of the two solutions (\ref{lfp13}) and (\ref{lfp18}). The solution (\ref{lfp13}) leads to a Dirac peak ($\dot\chi_+\neq 0$) contrary to  the solution (\ref{lfp18}) for which the normalization is conserved ($\dot\chi_-=0$). For $T>T'_*$ (i.e. $a<0$ or $\beta G m_1 m_2<d-2$), the singularity at $r=0$ behaves as a natural repulsive  boundary. In that case, the solution is unique and no boundary condition has to be fixed by hand. It is given by Eq. (\ref{lfp13}) which does not lead to a Dirac peak ($\dot\chi_+=0$).

\section{Conclusion}

In this paper, we have analytically studied the evolution of $N=2$
Brownian particles in gravitational interaction in a space of $d=2$
dimensions. Up to a change of notations, this is equivalent to the
simplified motion of two biological entities interacting via
chemotaxis (in which case the dimension $d=2$ is physically relevant).
Of course, the consideration of only $N=2$ particles is an extreme
limit but the problem is already involved and shows that the dynamics
is complex since the particles can coalesce to form Dirac peaks. The
same phenomenon (collapse and Dirac peaks) occurs for a larger number
of particles and has been investigated analytically in the mean field
limit $N\rightarrow +\infty$ \cite{herrerobio,sc,lushnikov}. It shares
some analogies with the Bose-Einstein condensation
\cite{sopik}. The case of a finite number of particles will be
investigated numerically in a forthcoming paper \cite{mc}.

Finally, we would like to point out some analogies with
the transport of passive particles by a stochastic turbulent flow
characterized by scale invariant structure functions (Kraichnan model)
\cite{gv,weinan,gh} or, more generally, with correlated Brownian
motions with scale invariant correlations \cite{gc}.  In particular,
implosive collapse of trajectories has been found by Gaw\c edzki \&
Vergassola \cite{gv} for strongly compressible flows. These authors
determined the statistics of inter-trajectory distances and observed a
lack of normalization when the diffusivity tends to zero. Like in our
problem, a defect of probability concentrates at $r=0$ in a
$\delta$-function term carrying the missing probability.  These
authors also studied the long time behavior of the averaged powers of
the distance between the Lagrangian trajectories and obtained power
law behaviors. In these hydrodynamical problems, the choice of the
boundary condition at $r=0$ when viscosity and diffusivity go to zero
is crucial and different regimes have been investigated.  In the case
of weak compressibility, the singularity at $r=0$ acts as a repulsive
entrance boundary and the particles never collide. For strong
compressibility and smooth flows, the point $r=0$ behaves as a natural
attractive boundary for which particles approach $r=0$ in an infinite
mean time. For strong compressibilities, $r=0$ works as an adhesive
boundary for which particles collide in finite time with a vanishing
relative velocity. In that case, they remain at that point indefinitely
(coalescence). Since the probability to find at $t>0$ the particle at
$r=0$ is finite and increases in time, no stationary state is reached
and $P(r,t)$ develops a Dirac delta function at $r=0$ with a time
increasing coefficient. For intermediate compressibilities, $r=0$
works as a regular boundary because particles hit one another in
finite time but with non-zero relative velocity. In that case, both
attractive and repulsive solutions are possible and it is necessary to
fix by hand a boundary condition at $r=0$. These different regimes
have been discussed in these terms by Gabrielli \& Cecconi \cite{gc}
in relation to boundary (or singularity) classification introduced by
Van Kampen \cite{vankampen} (see also Feller \cite{feller}). A similar 
phenomenology is obtained in the framerwork of our Brownian model.

\vskip0.4cm
{\it Acknowledgment}: We thank the referee for mentioning the
connection of our study with the transport of a passive particle by a
stochastic turbulent flow (Refs. \cite{gv,weinan,gh,gc}) and indicating
to us the van Kampen classification. This led to a more
detailed analysis of our model.

\appendix

\section{Moment of inertia and critical temperatures}
\label{sec_mict}

It is shown in \cite{epjb} (see also Appendix \ref{sec_virg}) that, in $d=2$, the total moment of inertia of self-gravitating Brownian particles
\begin{equation}
I_{tot}=\sum_{\alpha}\langle m_{\alpha}r_{\alpha}^2\rangle,
\label{mm1}
\end{equation}
satisfies the virial theorem\footnote{According to the discussion of Sec. \ref{sec_prob}, we now know that this relation ceases to be exact when the particles form Dirac peaks. However, we shall not address this problem here.}
\begin{equation}
\frac{1}{4}\xi \frac{d I_{tot}}{dt}=Nk_B(T-T_c)-PV,
\label{mm2}
\end{equation}
where $P$ is the total pressure at the boundary of the domain.  The
critical temperature appearing in this relation is
\begin{equation}
k_B T_c=\frac{G\sum_{\alpha\neq\beta}m_{\alpha}m_{\beta}}{4N}=\frac{G}{4N}\left (M^2-\sum_{\alpha=1}^{N}m_{\alpha}^2\right ).
\label{mm3}
\end{equation}
For equal mass particles
\begin{equation}
k_B T_c=(N-1)\frac{Gm^2}{4}.
\label{mm4}
\end{equation}

In fact, it may be more relevant to measure the positions of the particles relative to
the center of mass
\begin{equation}
{\bf R}=\frac{\sum_{\alpha}m_{\alpha}{\bf r}_{\alpha}}{M},
\label{mm5}
\end{equation}
and define the moment of inertia by
\begin{equation}
I=\sum_{\alpha}\langle m_{\alpha}({\bf r}_{\alpha}-{\bf R})^2\rangle.
\label{mm6}
\end{equation}
Indeed, if all the particles collapse in a single point, this point will be the center of mass ${\bf R}$. Therefore, $I$ will be zero while $I_{tot}$ is non zero. We clearly have the relation
\begin{equation}
I=I_{tot}-M\langle R^2\rangle.
\label{mm7}
\end{equation}
We also check that, for $N=2$, $I$ represents the moment of inertia of the reduced particle.  The center of mass has a pure Brownian motion
\begin{equation}
\frac{d{\bf R}}{dt}=\sqrt{2D_*}{\bf B}(t),  \label{mm8}
\end{equation}
with a diffusion coefficient
\begin{equation}
D_*=\frac{k_B T}{\xi M}.\label{mm9}
\end{equation}
It satisfies a relation of the form
\begin{equation}
\frac{1}{4}\xi M\frac{d\langle R^2\rangle}{dt}=k_B T-P_{eff}V,  \label{mm10}
\end{equation}
where $P_{eff}$ is an effective pressure on the boundary of the box
due to the center of mass.  Therefore, the virial theorem expressed in
terms of $I$ is
\begin{equation}
\frac{1}{4}\xi \frac{dI}{dt}=Nk_B(T-T_c)-k_B T-\Delta P V,
\label{mm11}
\end{equation}
where $\Delta P=P-P_{eff}$.  Since the motion of the center of mass is
completely decoupled, it is as if we had only $N-1$ particles in the
system. Therefore, it makes sense to rewrite the foregoing equation in
the form
\begin{equation}
\frac{1}{4}\xi \frac{dI}{dt}=(N-1) k_B(T-T_*)-\Delta P V,
\label{mm12}
\end{equation}
where
\begin{equation}
k_B T_*=\frac{N}{N-1}k_B T_c.
\label{mm13}
\end{equation}
For equal mass particles, we have
\begin{equation}
k_B T_*=N\frac{Gm^2}{4}.
\label{mm14}
\end{equation}
Furthermore, if we consider an infinite domain and measure the displacement of the particles relative to the center of mass, using Eq. (\ref{mm12}), we find that $\langle ({\bf r}-{\bf R})^2\rangle= 4D(T)t+\langle ({\bf r}-{\bf R})^2\rangle_0$ with an effective diffusion coefficient
\begin{equation}
D(T)=\frac{N-1}{N}\frac{k_{B}T}{\xi m}\left (1-\frac{T_*}{T}\right ).
\label{mm14bis}
\end{equation}
For $N=2$ particles, noting that $\mu=m/2$ and $\langle ({\bf r}-{\bf
R})^2\rangle=(\langle ({\bf r}_1-{\bf R})^2\rangle+\langle ({\bf
r}_2-{\bf R})^2\rangle)/2=\langle r^2\rangle/4$, 
where $\langle r^2\rangle$ denotes the
mean square displacement of the reduced particle, we find that
Eq. (\ref{mm14bis}) is consistent with Eq. (\ref{ew}).

Coming back to the general expression (\ref{mm12}), we conclude that
the system is expected to collapse and form Dirac peak(s) for $T<T_*$
and evaporate (in an infinite domain) or tend to an equilibrium state
(in a finite domain) for $T>T_*$. The temperature $T_{*}$ is precisely
the collapse temperature that was obtained in \cite{epjb} directly
from the study of the partition function in a bounded domain (the
partition function diverges for $T<T_*$). We now understand better the
relationship between the two temperatures $T_c$ and $T_*$. The
temperature $T_*$ is associated to the collapse of the system. It
arises in the virial theorem after the contribution of the center of
mass has been removed. However, the critical temperature appearing in
the exact equation of state is $T_c$. Indeed, at statistical
equilibrium, we have the relation \cite{epjb}:
\begin{equation}
PV=Nk_B(T-T_c),
\label{mm15}
\end{equation}
which also results from Eq. (\ref{mm2}). At $T=T_*$, the particles
form a Dirac containing all the mass and the equation of state
(\ref{mm15}) reduces to
\begin{equation}
PV=k_B T_*,
\label{mm15b}
\end{equation}
where we have used Eq. (\ref{mm13}). This is the equation of state of a  single Brownian particle (the Dirac) at temperature $T=T_*$. It is of the usual form $PV=Nk_BT$ with $N=1$.

\section{Check of consistency for $T\neq 0$}
\label{sec_c1}

In this Appendix, we check that the relation (\ref{m3}) is consistent
with the virial theorem (\ref{fp10}). The virial theorem (\ref{fp10}) can be rewritten
\begin{equation}
\dot I=4D\mu \chi(t)(1-a).
\label{ca1}
\end{equation}
Integrating this relation, we get
\begin{equation}
I(t)=4D(1-a)\mu\int_{0}^{t}\chi(\tau)\, d\tau+I(0),
\label{ca2}
\end{equation}
with $I(0)=Mr_0^2$ in our case. Using Eq. (\ref{d1}), we have
\begin{equation}
\int_{0}^{t}\chi(\tau)\, d\tau=t-\int_{0}^{t}\frac{\Gamma_a\left (\frac{r_0^2}{4D\tau}\right )}{\Gamma(a)}\, d\tau.
\label{ca3}
\end{equation}
Setting $x=r_0^2/(4D\tau)$, this can be rewritten
\begin{equation}
\int_{0}^{t}\chi(\tau)\, d\tau=t-\frac{1}{\Gamma(a)}\frac{r_0^2}{4D}\int_{\frac{r_0^2}{4Dt}}^{+\infty}\Gamma_a(x)\frac{dx}{x^2}.
\label{ca4}
\end{equation}
Let us consider the integral
\begin{equation}
K(s)=\int_{s}^{+\infty}\frac{\Gamma_a(x)}{x^2}\, dx,
\label{ca5}
\end{equation}
where we recall that
\begin{equation}
\Gamma_a(x)=\int_{x}^{+\infty}t^{a-1}e^{-t}\, dt.
\label{ca6}
\end{equation}
Integrating by parts, we get
\begin{equation}
K(s)=\frac{\Gamma_a(s)}{s}-\int_{s}^{+\infty}x^{a-2}e^{-x}\, dx.
\label{ca7}
\end{equation}
Integrating by parts again, we obtain
\begin{equation}
K(s)=\left (\frac{1}{s}+\frac{1}{1-a}\right )\Gamma_a(s)-\frac{s^{a-1}e^{-s}}{1-a}.
\label{ca8}
\end{equation}
We now have
\begin{equation}
I(t)=\mu r_0^2+4D(1-a)\mu\left \lbrack t-\frac{1}{\Gamma(a)}\frac{r_0^2}{4D}K\left (\frac{r_0^2}{4Dt}\right )\right \rbrack.
\label{ca9}
\end{equation}
Substituting Eq. (\ref{ca8}) in (\ref{ca9}), we recover Eq. (\ref{m3}).

\section{The case $T=0$}
\label{sec_z}

The case $T=0$ can be treated specifically\footnote{This study can be easily generalized in $d$ dimensions.}. In that case, the reduced particle has a deterministic motion given by the equation
\begin{equation}
\frac{d{\bf r}}{dt}=-k\frac{\bf r}{r^2},  \label{z1}
\end{equation}
where
\begin{equation}
k=\frac{Gm_1m_2}{\xi\mu}.  \label{z2}
\end{equation}
This equation can be integrated at once. If $a$ denotes the initial position of the particle, its position at time $t$ is
\begin{equation}
r^2=a^2-2kt.  \label{z3}
\end{equation}
Therefore, the particle reaches the origin at a time
\begin{equation}
t(a)=\frac{a^2}{2k}. \label{z4}
\end{equation}
Equivalently, the particle that reaches the origin at time $t$ was located initially at
\begin{equation}
a(t)=\sqrt{2kt}. \label{z4b}
\end{equation}
Let $P_0(a)$ be the initial probability density to find the particle in $a$. The conservation of the probability density imposes
\begin{equation}
P(r,t)2\pi r dr=P_0(a)2\pi a da. \label{z5}
\end{equation}
Now, according to the equation of motion (\ref{z3}), we have $r dr= a da$ so that $P(r,t)=P_0(a)$. Therefore, the probability density to find the particle in $r$ at time t is
\begin{equation}
P(r,t)=P_0 \left (\sqrt{r^2+2kt}\right ). \label{z6}
\end{equation}
We can check by direct substitution that this is indeed the solution of the Fokker-Planck equation (\ref{fp1}) at $T=0$:
\begin{equation}
{\partial P\over \partial t}={k\over r}{\partial P\over\partial r}. \label{z7}
\end{equation}
The probability that the particle has not reached $r=0$ at time $t$ is
\begin{eqnarray}
\chi(t)=\int_0^{+\infty}P(r,t)2\pi r\, dr.
 \label{z8}
\end{eqnarray}
Using the distribution (\ref{z6}) and performing the change of variables (\ref{z3}), we get
\begin{eqnarray}
\chi(t)=\int_{\sqrt{2kt}}^{+\infty}P_0(a)2\pi a\, da
=1-\int_{0}^{\sqrt{2kt}}P_0(a)2\pi a\, da. \nonumber\\
 \label{z9}
\end{eqnarray}
Therefore, the probability that the particle has formed a Dirac peak at time $t$ is
\begin{eqnarray}
\chi_D(t)=\int_{0}^{\sqrt{2kt}}P_0(a)2\pi a\, da.
 \label{z10}
\end{eqnarray}
This corresponds to probability to find initially the particle in the disk of radius $\sqrt{2kt}$. For consistency, let us derive this result in a different manner. Starting from the relation
\begin{eqnarray}
\dot \chi_D=2\pi k P(0,t),
 \label{z11}
\end{eqnarray}
and using Eq. (\ref{z6}), we get
\begin{eqnarray}
\dot \chi_D=2\pi k P_0 \left (\sqrt{2kt}\right ).
 \label{z12}
\end{eqnarray}
Integrating this relation, we find that
\begin{eqnarray}
\chi_D(t)=2\pi k\int_{0}^{t} P_0 \left (\sqrt{2k\tau}\right )\, d\tau
=\int_{0}^{\sqrt{2kt}}P_0(a)2\pi a\, da.\nonumber\\
 \label{z13}
\end{eqnarray}
This returns Eq. (\ref{z10}) as it should. Finally, the moment of inertia is
\begin{eqnarray}
I(t)=\int_{0}^{+\infty}P(r,t)\mu r^2 2\pi r\, dr.
 \label{z14}
\end{eqnarray}
Substituting the probability density (\ref{z6}) in Eq. (\ref{z14}) and
performing the change of variables (\ref{z3}), we obtain
\begin{eqnarray}
I(t)=\int_{\sqrt{2kt}}^{+\infty} P_0(a) \mu (a^2-2kt) 2\pi a\, da.
 \label{z15}
\end{eqnarray}
This can be written equivalently
\begin{eqnarray}
I(t)=\int_{\sqrt{2kt}}^{+\infty} P_0(a)\mu a^2 2\pi a\, da-2k\mu \chi(t)t.
 \label{z16}
\end{eqnarray}
In Appendix \ref{sec_c2}, we check that this relation is consistent with the virial theorem (\ref{fp10}).

Let us specifically apply these results to an initial 
distribution of the form (\ref{lp2}). Since the motion is
deterministic, the particle initially located at $r_0$ will be located
at $r(t)=\sqrt{r_0^2-2kt}$ at time $t$. It will reach the origin in a
{\it finite} time $t_c=r_0^2/(2k)$ (at that time, the two original
particles stick together and remain tightly
bound). Therefore, the probability
that the particle has formed a Dirac at time $t$ is a Heaviside
function: $\chi_{D}=0$ if $t<t_c$ and $\chi_{D}=1$ if $t>t_c$. The
moment of inertia is $I(t)=\mu r(t)^2$ if $t\le t_c$ and $I(0)=0$ if
$t\ge t_c$.

\section{Check of consistency for $T=0$}
\label{sec_c2}

For $T=0$, the virial theorem (\ref{fp10}) becomes
\begin{equation}
\dot I=-2k\mu \chi(t).
\label{cb1}
\end{equation}
Integrating this relation, we get
\begin{equation}
I(t)=I(0)-2k\mu \int_{0}^{t}\chi(\tau)\, d\tau.
\label{cb2}
\end{equation}
Integrating by parts, we have
\begin{equation}
\int_{0}^{t}\chi(\tau)\, d\tau=t\chi(t)-\int_{0}^{t}\dot \chi(\tau)\tau\, d\tau.
\label{cb3}
\end{equation}
According to Eqs. (\ref{fp6}) and (\ref{z6}), we have
\begin{equation}
\dot \chi(\tau)=-2\pi k P_{0}(\sqrt{2kt}).
\label{cb4}
\end{equation}
Therefore
\begin{equation}
I(t)=I(0)-2k\mu \chi(t)t-4\pi k^2\mu \int_{0}^{t}P_0(\sqrt{2k\tau})\tau\, d\tau.
\label{cb5}
\end{equation}
Setting $a=\sqrt{2k\tau}$, we obtain
\begin{equation}
I(t)=I(0)-2k\mu\chi(t)t-2\pi\mu \int_{0}^{\sqrt{2kt}}P_0(a) a^3\, da.
\label{cb6}
\end{equation}
Since
\begin{equation}
I(0)=2\pi\mu \int_{0}^{+\infty}P_0(a) a^3\, da,
\label{cb7}
\end{equation}
we finally recover Eq. (\ref{z16}).

\section{Asymptotic behaviors}
\label{sec_asym}

In this Appendix, we determine the asymptotic behaviors of the function
\begin{eqnarray}
f(x)=\frac{1}{\lbrack x\Gamma(x)\rbrack^{1/x}},
\label{as1}
\end{eqnarray}
appearing in Eq. (\ref{d6b}). We first determine the behavior of the Gamma function for small $x$. Expanding the identity
\begin{eqnarray}
\Gamma(1-x)\Gamma(x)=\frac{\pi}{\sin(\pi x)},
\label{as2}
\end{eqnarray}
for $x\rightarrow 0$, and using
\begin{eqnarray}
\Gamma'(1)=-\gamma\simeq -0.57721...,
\label{as3}
\end{eqnarray}
where $\gamma$ is the Euler constant, we obtain
\begin{eqnarray}
x\Gamma(x)\sim 1-\gamma x, \qquad (x\rightarrow 0).
\label{as4}
\end{eqnarray}
From the previous result, we deduce that
\begin{eqnarray}
\ln\left\lbrace \lbrack x\Gamma(x)\rbrack^{1/x}\right\rbrace\sim \frac{1}{x}\ln (1-\gamma x)\rightarrow  -\gamma.
\label{as5}
\end{eqnarray}
Therefore, for $x\rightarrow 0$, we find that
\begin{eqnarray}
\frac{1}{\lbrack x\Gamma(x)\rbrack^{1/x}}\rightarrow e^{\gamma}\simeq 1.78107...
\label{as6}
\end{eqnarray}
On the other hand, using the equivalent
\begin{eqnarray}
\Gamma(x)\sim \sqrt{2\pi} e^{-x}x^{x-\frac{1}{2}},\qquad (x\rightarrow +\infty),
\label{as7}
\end{eqnarray}
we find, for $x\rightarrow +\infty$, that
\begin{eqnarray}
\frac{1}{\lbrack x\Gamma(x)\rbrack^{1/x}}\sim \frac{e}{x}.
\label{as8}
\end{eqnarray}

Considering now the function defined by Eq. (\ref{d4b}) and using the equivalent
\begin{eqnarray}
\Gamma(x)\sim \frac{1}{x},\qquad (x\rightarrow 0),
\label{as9}
\end{eqnarray}
we find, for $a\rightarrow 0$, that
\begin{eqnarray}
\chi_{D}(1)\sim a\Gamma_0(1)\sim 0.219384 a.
\label{as10}
\end{eqnarray}
To determine its asymptotic behavior for $a\rightarrow +\infty$, we first note that $\Gamma_{a}(x)=\Gamma(a)-\gamma(a,x)$ where
\begin{eqnarray}
\gamma(a,x)=\int_{0}^{x}e^{-t}t^{a-1}\, dt.
\label{as11}
\end{eqnarray}
For $a\rightarrow +\infty$, we have the asymptotic expansion
\begin{eqnarray}
\gamma(a,x)\sim \sum_{n=0}^{+\infty}\frac{(-1)^n x^{a+n}}{(a+n)n!}.
\label{as12}
\end{eqnarray}
This implies $\gamma(a,1)\sim 1/(ae)$. On the other hand,
\begin{eqnarray}
\Gamma(a)\sim \sqrt{2\pi} e^{-a}a^{a-\frac{1}{2}}.
\label{as13}
\end{eqnarray}
Combining these results and recalling that $\chi_{D}(1)=1-\gamma(a,1)/\Gamma(a)$, we obtain
\begin{eqnarray}
1-\chi_{D}(1)\sim\frac{1}{\sqrt{2\pi}}\frac{e^{a-1}}{a^{a+1/2}},
\label{as14}
\end{eqnarray} 
for $a\rightarrow +\infty$.

\section{The van Kampen classification}
\label{sec_vk}

In this Appendix, we apply to our system the boundary classification
introduced by van Kampen \cite{vankampen}. For a summary, we refer to
Gabrielli and Cecconi \cite{gc}. In order to avoid repetitions, we
shall use their notations and we refer the reader to their paper for
more details.

For the sake of generality, we consider the logarithmic Fokker-Planck
equation in $d$ dimensions
\begin{eqnarray}
\frac{\partial P}{\partial t}=\frac{1}{r^{d-1}}\frac{\partial}{\partial r}\left\lbrace Dr^{d-1}\left (\frac{\partial P}{\partial r}+P\frac{\beta G m_1 m_2}{r}\right )\right\rbrace.
\label{vk1}
\end{eqnarray}
To apply the van Kampen classification, we
must first transform the Fokker-Planck equation (\ref{vk1}) into a one
dimensional Fokker-Planck equation. To that purpose, we set $r=\sqrt{D}x$ and
$f(x,t)=\sqrt{D}P(r,t)S_d r^{d-1}$. This transforms Eq. (\ref{vk1}) into
\begin{eqnarray}
\frac{\partial f}{\partial t}=\frac{\partial}{\partial x}\left (\frac{\partial f}{\partial x}-\frac{d-1}{x} f+\frac{\beta G m_1 m_2}{x}f\right ),
\label{vk2}
\end{eqnarray}
with the normalization condition $\int_{0}^{+\infty}f(x,t)\, dx=1$. This is a one dimensional Fokker-Planck equation of the form
\begin{eqnarray}
\frac{\partial f}{\partial t}=\frac{1}{2}\frac{\partial^2}{\partial x^2}\left (D(x)f\right )-\frac{\partial}{\partial x}\left (K(x)f\right ),
\label{vk3}
\end{eqnarray}
with $D(x)=2$ and $K(x)=(d-1-\beta Gm_1m_2)/x$. Van Kampen's classification for a singularity $x=0$ is based on the analysis of the behavior for $\epsilon\rightarrow 0$ of the integrals
\begin{eqnarray}
L_1=\int_{\epsilon}^{x_0}dx\, e^{\phi(x)},
\label{vk4}
\end{eqnarray}
\begin{eqnarray}
L_2=\int_{\epsilon}^{x_0}dx\, e^{\phi(x)}\int_{x_0}^{x}dx'\, \frac{e^{-\phi(x')}}{D(x')},
\label{vk5}
\end{eqnarray}
\begin{eqnarray}
L_3=\int_{\epsilon}^{x_0}dx\, \frac{e^{-\phi(x)}}{D(x)},
\label{vk6}
\end{eqnarray}
where
\begin{eqnarray}
\phi(x)=-2\int_{x_0}^{x}dx'\, \frac{K(x')}{D(x')},
\label{vk7}
\end{eqnarray}
and with $x_0>0$. For the logarithmic Fokker-Planck equation (\ref{vk2}), we have $D(x)=2$ and
\begin{eqnarray}
\phi(x)=-(d-1-\beta G m_1 m_2)\ln\left (\frac{x}{x_0}\right ).
\label{vk8}
\end{eqnarray}
Considering the limit $\epsilon\rightarrow 0$, it is easy to see that: (i) $L_1<+\infty$ iff $\beta Gm_1m_2>d-2$, (ii)  $L_2<+\infty$ iff $\beta Gm_1m_2>d-2$, (iii) $L_3<+\infty$ iff $\beta Gm_1m_2<d$.
Therefore, according to van Kampen's classification, we need to consider three cases (it is useful to introduce the critical temperatures $T_*=Gm_1m_2/d$ and $T_*'=Gm_1m_2/(d-2)$):

(i) if $\beta Gm_1m_2<d-2$, i.e. $T>T'_*$, the singularity $x=0$ behaves as a {\it natural repulsive boundary} ($L_1\rightarrow +\infty$). The particle run away from the singularity never touching it. The solution is unique.

(ii) if $d-2<\beta Gm_1m_2<d$, i.e. $T_*<T<T'_*$, the singularity $x=0$ behaves as a {\it regular boundary} ($L_1, L_2, L_3< +\infty$). In that case, an absorbing or reflecting boundary condition has to be fixed by hand to determine the solution of the equation.

(iii) if $\beta Gm_1m_2>d$, i.e. $T<T_*$, the singularity $x=0$ behaves as an {\it attractive adhesive boundary} ($L_1, L_2 < +\infty$ and $L_3\rightarrow +\infty$). In that case, $f(x,t)$ develops a Dirac peak at $x=0$ with a time increasing coefficient. The solution is unique.

For $d\le 2$, we only have a transition at temperature
$T_*=Gm_1m_2/d$. For $d>2$, we have two transitions at temperatures
$T_*=Gm_1m_2/d$ and $T_*'=Gm_1m_2/(d-2)$.

\section{Temporal correlation functions and front structure of the logarithmic Fokker-Planck equation}
\label{sec_corr}

We consider the logarithmic Fokker-Planck equation (\ref{vk1}) in a space of dimension $d$. We assume that the domain is unbounded. In order to have an equilibrium state $P_e({\bf r})$, the potential must be regularized at short distances. Therefore, we assume that the potential has a logarithmic behavior for sufficiently large $r$ and that it tends to a finite constant for $r\rightarrow 0$.  In that case, $P_e\propto r^{-\beta G m_1 m_2}$ for $r\rightarrow +\infty$, and there exists an equilibrium state iff $k_B T<k_B T_*=G m_1 m_2/d$. To determine the temporal correlation functions, we use the theory of Marksteiner {\it et al.} \cite{mark} (see also \cite{farago,lillo,lutz,bd,clvortex,new}). As shown in Appendix \ref{sec_vk}, the Fokker-Planck equation (\ref{vk1}) can be transformed into a one dimensional Fokker-Planck equation of the form
\begin{eqnarray}
\frac{\partial f}{\partial t}=\frac{\partial}{\partial x}\left (\frac{\partial f}{\partial x}+f\frac{\partial\Phi}{\partial x}\right ),
\label{corr1}
\end{eqnarray}
with a potential behaving like $\Phi(x)\sim\alpha \ln x$ for $x\rightarrow +\infty$, where $\alpha=\beta G m_1 m_2 -(d-1)$. We introduce the temporal correlation functions  $C(t)=\langle A(0)A(t)\rangle-\langle A\rangle^2$ and refer to \cite{mark,farago,lillo,lutz,bd,clvortex,new} for the details of the calculations (we use here the notations of \cite{clvortex}). If $A(t)=x(t)^n$, then
\begin{eqnarray}
C(t)\sim t^{-\xi},\quad \xi=-n+\frac{\alpha-1}{2}.
\label{corr2}
\end{eqnarray}
In the present case, the exponent can be written $\xi=-n+(\beta G m_1 m_2 -d)/2$. Equation (\ref{corr2}) is valid provided that $\xi>0$, i.e. $k_B T<Gm_1m_2/(2n+d)$, which corresponds to the condition of existence of the moment $\langle x^{2n}\rangle$ at equilibrium. If $A(t)\sim (\ln x(t))^{1/\delta}$, then
\begin{eqnarray}
C(t)\sim \frac{(\ln t)^{2/\delta}}{t^{\frac{\alpha-1}{2}}}.
\label{corr3}
\end{eqnarray}

We note, finally, that the relaxation of the tail of the distribution function $P(r,t)$ that is solution of the logarithmic Fokker-Planck equation (\ref{vk1})  can be studied with the approach developed by Chavanis \& Lemou \cite{lemoufront}. In particular, the function $u(r,t)=P(r,t)/P_e(r)$  has a front structure and the position of the front evolves in time like $r_f(t)\sim \sqrt{2D\alpha t}$.

\section{Virial theorem in the post-collapse regime of the Smoluchowski-Poisson system}
\label{sec_post}

A virial theorem associated with the Smoluchowski-Poisson system in
$d=2$ dimensions has been derived in \cite{virial}. However, the
derived equation is not valid in the post-collapse dynamics when a
Dirac peak forms at ${\bf r}={\bf 0}$ and grows. This happens for
$T<T_c=GMm/(4k_B)$ when $t>t_{coll}$ \cite{virial}. The reason is the
same as the one given in Sec. \ref{sec_fp}. In this Appendix, we
provide the proper form of the virial theorem that is valid both in
the pre and post collapse regimes.

The total density profile of the self-gravitating Brownian gas can be written
\begin{equation}
\rho({\bf r},t)=M_D(t)\delta({\bf r})+\rho({\bf r},t),
\label{post1}
\end{equation}
where the first term takes into account the possible formation of a
Dirac peak at ${\bf r}={\bf 0}$ and the second term is the (regular)
density profile excluding the Dirac. The total mass is $M=M_D(t)+M(t)$ where $M_D(t)$ is the mass contained in the Dirac peak and $M(t)=\int\rho({\bf r},t)\, d{\bf r}$ is the mass outside the Dirac. The Smoluchowski-Poisson system accounting for the presence of a Dirac peak can be written 
\begin{equation}
\xi\frac{\partial\rho}{\partial t}=\nabla\cdot \left (\frac{k_B T}{m}\nabla\rho+\rho\frac{GM_D(t)}{r^{d}}{\bf r}+\rho\nabla\Phi\right ),
\label{post2}
\end{equation}
\begin{equation}
\Delta\Phi=S_d G\rho.
\label{post3}
\end{equation}
Using $\dot M_D=-\dot M$ and integrating Eq. (\ref{post2}) on the infinite space, we find that the mass accumulated in the Dirac peak by unit of time is
\begin{equation}
\frac{dM_D}{dt}=\frac{S_d G}{\xi}M_D(t)\rho(0,t).
\label{post4}
\end{equation}
Equations (\ref{post2})-(\ref{post4}) form a closed system describing the evolution of the system in the pre and post collapse regimes. These equations have been studied in \cite{post}.

We now specialize on the 2D Smoluchowski-Poisson system. Taking the time derivative of the moment of inertia
\begin{equation}
I=\int\rho\, d{\bf r},
\label{post5}
\end{equation} 
and using Eq. (\ref{post2}), we obtain after integrations by parts
\begin{equation}
\xi\frac{dI}{dt}=\frac{4 k_B T}{m}M(t)-2GM(t)M_{D}(t)+2W_{ii},
\label{post6}
\end{equation} 
where $W_{ii}=-\int\rho {\bf r}\cdot\nabla\Phi\, d{\bf r}$ is the virial of the gravitational force. In $d=2$ dimensions, it is equal to $W_{ii}=-GM(t)^2/2$ (see \cite{virial} and Appendix \ref{sec_virg}). Therefore, we obtain the virial theorem
\begin{equation}
\frac{1}{4}\xi\frac{dI}{dt}=M(t)\left ( \frac{k_B T}{m}-\frac{GM_{D}(t)}{2}-\frac{GM(t)}{4}\right ),
\label{post7}
\end{equation} 
that is valid in all the regimes of the dynamics. For $T>T_c$ or in the pre-collapse regime $t<t_{coll}$ for $T<T_c$, there is no Dirac peak at ${\bf r}={\bf 0}$. In that case, $M_{D}(t)=0$, $M(t)=M$, and the virial theorem (\ref{post7}) reduces to
\begin{equation}
\frac{1}{4}\xi\frac{dI}{dt}=Nk_B (T-T_c),
\label{post8}
\end{equation}  
where $k_BT_c=GMm/4$. This returns the result of \cite{virial}. Let us now consider the post collapse regime $t>t_{coll}$ for $T<T_c$. When $t\rightarrow +\infty$, almost all the mass is in the Dirac so that $M_{D}(t)\simeq M$ and $M(t)=M-M_{D}(t)\simeq 0$. In that case, the Smoluchowski equation (\ref{post2}) can be approximated by 
\begin{equation}
\xi\frac{\partial\rho}{\partial t}=\nabla\cdot \left (\frac{k_B T}{m}\nabla\rho+\rho\frac{GM}{r^{2}}{\bf r}\right ),
\label{post9}
\end{equation}
and Eq. (\ref{post4}) becomes
\begin{equation}
\frac{dM_D}{dt}=\frac{S_d G}{\xi}M\rho(0,t).
\label{post10}
\end{equation}
This is equivalent to Eqs. (\ref{fp1}) and (\ref{fp7}) studied in this paper, with a simple change of notations discussed in Sec. \ref{sec_bp}. In that case, the virial theorem (\ref{post7}) becomes
\begin{equation}
\frac{1}{4}\xi\frac{dI}{dt}=N(t)k_B (T-T_*),
\label{post11}
\end{equation}  
where $k_BT_*=GMm/2$ (i.e. $T_*=2T_c$).  This is equivalent to the virial theorem (\ref{fp10}).

\section{Virial theorem for power-law interactions}
\label{sec_virg}

In this Appendix, we provide the general form of the virial theorem
for Brownian particles with power law interactions in $d$
dimensions. We only give the
final expressions, and refer to \cite{virial} for more details on
their derivation. As explained in Sec. \ref{sec_prob}, the following
expressions are valid as long as there are no Dirac peaks.

Let us consider $N$ Brownian particles with individual mass $m_{\alpha}$ in a space of dimension $d$. We assume that the particles are subject to an external harmonic potential $V({\bf r})=\frac{1}{2}\omega_0^2 r^2$ and that they interact through an algebraic potential $u(\xi)=-\frac{1}{d+\gamma-2}\frac{G}{\xi^{d+\gamma-2}}$ if $\gamma\neq 2-d$ and through a logarithmic potential $u(\xi)=G\ln\xi$ if $\gamma=2-d$, both corresponding to a force $-u'(\xi)=-G/\xi^{d+\gamma-1}$. The gravitational potential is recovered for $\gamma=0$. The case studied in the present paper is very particular because it corresponds to a logarithmic ($\gamma=2-d$) {\it and} a Newtonian ($\gamma=0$) interaction. The stochastic equations of motion of the particles are
\begin{equation}
\ddot x_{i}^{\alpha}=\sum_{\beta\neq \alpha}
{Gm_{\beta}(x_{i}^{\beta}-x_{i}^{\alpha})\over |{\bf
r}_{\beta}-{\bf r}_{\alpha}|^{d+\gamma}}-\omega_0^2 x_{i}^{\alpha}-\xi \dot
x_{i}^{\alpha}+\sqrt{2D_{\alpha}}B_{i}^{\alpha}(t), \label{virg1}
\end{equation}
where ${\bf B}_{\alpha}(t)$ is a white noise. Here, the Greek letters
refer to the particles and the Latin letters to the coordinates of
space.  The diffusion coefficient is given by the Einstein formula
$D_{\alpha}=\xi k_{B}T/m_{\alpha}$. The moment of inertia tensor is
defined by
\begin{equation}
I_{ij}=\sum_{\alpha}m_{\alpha}x_{i}^{\alpha}x_{j}^{\alpha}.
\label{virg2}
\end{equation}
We introduce the kinetic energy tensor
\begin{equation}
K_{ij}={1\over 2}\sum_{\alpha}m_{\alpha}{\dot x}_{i}^{\alpha}{\dot x}_{j}^{\alpha},
\label{virg3}
\end{equation}
and the potential energy tensor
\begin{eqnarray}
W_{ij}=G\sum_{\alpha\neq\beta} m_{\alpha}m_{\beta}{x_{i}^{\alpha}
(x_{j}^{\beta}-x_{j}^{\alpha})\over |{\bf r}_{\beta}
-{\bf r}_{\alpha}|^{d+\gamma}}\nonumber\\
=-{1\over 2}G \sum_{\alpha\neq\beta}
m_{\alpha}m_{\beta}{(x_{i}^{\alpha}-x_{i}^{\beta})
(x_{j}^{\alpha}-x_{j}^{\beta})\over
|{\bf r}_{\beta}-{\bf r}_{\alpha}|^{d+\gamma}}, \label{virg4}
\end{eqnarray}
where the second equality results from simple algebraic manipulations
obtained by interchanging the dummy variables $\alpha$ and $\beta$ and summing the resulting expressions.
The tensor virial theorem associated with the stochastic equations (\ref{virg1}) is
\begin{eqnarray}
{1\over 2} {\ddot I}_{ij}+{1\over 2}\xi  {\dot
I}_{ij}+\omega_0^2 I_{ij} =2K_{ij}+W_{ij} \nonumber\\
-{1\over 2}\oint
(P_{ik}x_{j}+P_{jk}x_{i})\,dS_{k}, 
\label{virg5}
\end{eqnarray}
where the last term takes into account pressure forces at the boundary of the system.
For Brownian particles, it is implicitly assumed that the quantities appearing in Eq. (\ref{virg5}) are
averaged over the noise and over statistical realizations, while for Hamiltonian systems ($\xi=D_\alpha=0$), Eq. (\ref{virg5}) is exact without averages.  The scalar virial
theorem is obtained by contracting the indices leading to
\begin{eqnarray}
{1\over 2} {\ddot I} +{1\over 2}\xi  {\dot I} +\omega_0^2 I=2 K+ W_{ii} -\oint P_{ik}x_{i}dS_{k},
\label{virg6}
\end{eqnarray}
where
\begin{eqnarray}
I=\sum_{\alpha}m_{\alpha}x_{\alpha}^{2},\qquad K={1\over 2}
\sum_{\alpha}m_{\alpha}v_{\alpha}^{2},
\label{virg7}
\end{eqnarray}
are the moment of inertia and the kinetic energy. On the other hand, $W_{ii}$ is the virial which takes the form
\begin{eqnarray}
W_{ii}=-{1\over 2}G\sum_{\alpha\neq\beta} {m_{\alpha}m_{\beta}\over
|{\bf r}_{\beta}-{\bf r}_{\alpha}|^{d+\gamma-2}}.
\label{virg8}
\end{eqnarray}

For $\gamma\neq 2-d$, we find that
\begin{eqnarray}
W_{ii}=(d+\gamma-2)W,
\label{virg9}
\end{eqnarray}
where $W$ is the potential energy
\begin{eqnarray}
W=-{G\over 2(d+\gamma-2)}\sum_{\alpha\neq\beta}{m_{\alpha}m_{\beta}\over
|{\bf r}_{\beta}-{\bf r}_{\alpha}|^{d+\gamma-2}}. \label{virg10}
\end{eqnarray}
In that case, the scalar virial theorem reads
\begin{eqnarray}
{1\over 2} {\ddot I} +{1\over 2}\xi  {\dot I}+\omega_0^2 I =2 K +(d+\gamma-2)W -\oint P_{ik}x_{i}dS_{k}.\nonumber\\
\label{virg11}
\end{eqnarray}
For Hamiltonian systems ($D=\xi=0$), the scalar virial theorem in an unbounded domain  ($P=0$) reduces to \cite{chandravirial,bt}:
\begin{eqnarray}
{1\over 2} {\ddot I}+\omega_0^2 I =2 K +(d+\gamma-2)W.
\label{virg12}
\end{eqnarray}
Since the total energy $E=K+W+\frac{1}{2}\omega_0^2 I$ is
conserved, we obtain
\begin{eqnarray}
{1\over 2} {\ddot I}+2\omega_0^2 I=2E+(d+\gamma-4)W.
\label{virg13}
\end{eqnarray}
For the index $\gamma=4-d$, we get 
\begin{eqnarray}
{\ddot I}+4\omega_0^2 I=4E.
\label{virg14}
\end{eqnarray}
When $\omega_0^2=0$, we obtain ${\ddot I}=4E$ which yields after
integration $I=2Et^2+C_1 t+C_2$. For $E>0$, $I\rightarrow +\infty$
indicating that the system evaporates. For $E<0$, $I$ goes to zero in
a finite time, indicating that the system forms a Dirac peak in a
finite time.  When $\omega_0^2> 0$, the moment of inertia $I$
oscillates with pulsation $2\omega_0$ around the value
$E/\omega_0^2$. For $\omega_0^2=-\Omega_0^2<0$, corresponding to a
repulsive harmonic potential (or a rotation), the moment of inertia
increases exponentially rapidly as $e^{2\Omega_0 t}$. For the
gravitational interaction ($\gamma=0$), the relation (\ref{virg14}) is
valid in a space of dimension $d=4$ \cite{virial}. For $d=1$, this
relation is valid for the potential
$u=-\frac{1}{2}\frac{G}{\xi^2}$. This is related to the
Calogero-Sutherland model \cite{calogero,sutherland}.

For $\gamma=2-d$, corresponding to a logarithmic potential in $d$ dimensions, we have the simple exact result
\begin{eqnarray}
W_{ii}=-{1\over 2}G\sum_{\alpha\neq\beta}m_{\alpha}m_{\beta}.
\label{virg15}
\end{eqnarray}
It is interesting to note that this expression only depends on the mass of the particles and not on their position. For equal mass particles,
\begin{eqnarray}
W_{ii}=-{1\over 2}G N(N-1)m^{2}.
\label{virg16}
\end{eqnarray}
Since
\begin{eqnarray}
\sum_{\alpha\neq\beta}m_{\alpha}m_{\beta}=M^{2}-\sum_{\alpha}m_{\alpha}^{2},
\label{virg17}
\end{eqnarray}
we see that the first term is of order $N^{2} \overline{m}^{2}$ and the second of order $N
\overline{m}^{2}$ (where $\overline{m}$ is a typical mass). Therefore, in the mean-field
limit $N\rightarrow +\infty$, we obtain
\begin{eqnarray}
W_{ii}^{m.f.}=-{G M^{2}\over 2},
\label{virg18}
\end{eqnarray}
whatever the number of species in the system.

At equilibrium, the scalar virial theorem (\ref{virg6}) reduces to
\begin{eqnarray}
2 K +W_{ii}-\omega_0^2 I =\oint P_{ik}x_{i}dS_{k}.
\label{virg20}
\end{eqnarray}
For Hamiltonian systems, this relation is valid for a steady state
after time averages, or averages over statistical realizations, have
been made. If the system is at statistical equilibrium, then $K
={d\over 2}Nk_{B}T$ and $P_{ij}=p\delta_{ij}$ with
$p=\sum_{s}\rho_{s}k_{B}T/m_{s}$, where $\rho_s$ refers to the density
of the different species. Introducing the notation
$P=\frac{1}{dV}\oint p{\bf r}\cdot d{\bf S}$ \cite{virial}, we get
\begin{eqnarray}
dNk_{B}T +W_{ii}-\omega_0^2 I =dPV.
\label{virg21}
\end{eqnarray}
For an ideal gas without interaction $(W_{ii}=0$), we recover the
perfect gas law $PV+\omega_0^2 I/d=Nk_{B}T$ in the presence of a
harmonic potential (when $\omega_0^2>0$, we get in an unbounded domain
$I=dNk_{B}T/\omega_0^2$,  and when $\omega_0=0$, we get
$PV=Nk_{B}T$). Alternatively, for a gas with logarithmic interactions
($\gamma=2-d$), using Eq. (\ref{virg15}), we obtain the exact equation
of state
\begin{equation}
PV+\frac{\omega_0^2 I}{d}=Nk_{B}(T-T_{c}),
\label{virg22}
\end{equation}
with the {\it exact} critical temperature
\begin{equation}
k_{B}T_{c}={G \sum_{\alpha\neq\beta}m_{\alpha}m_{\beta}\over 2dN}.
\label{virg23}
\end{equation}
For equal mass particles, we get
\begin{equation}
k_{B}T_{c}=(N-1){Gm^{2}\over 2d}.
\label{virg24}
\end{equation}
In the mean-field limit
\begin{equation}
k_{B}T_{c}^{m.f.}={GM^{2}\over 2dN}.
\label{virg25}
\end{equation}
If $\omega_0= 0$, the equation of state (\ref{virg22}) reduces to
\begin{equation}
PV=Nk_{B}(T-T_{c}).
\label{virg22b}
\end{equation}
When $\omega_0^2\ge 0$, according to Eq. (\ref{virg22}), an
equilibrium state can possibly exist only for $T\ge T_c$ (since $P\ge
0$ and $I\ge 0$). For $T=T_c$, we have $P=0$ if $\omega_0^2=0$,
$P=I=0$ if $\omega_0^2>0$ and $I=dPV/\Omega_0^2$ if
$\omega_0^2=-\Omega_0^2<0$. In an unbounded domain ($P=0$), we get
\begin{equation}
\frac{\omega_0^2 I}{d}=Nk_{B}(T-T_{c}).
\label{virg26}
\end{equation}
If $\omega_0=0$, an equilibrium state can possibly exist only for
$T=T_c$. If $\omega_0^2=-\Omega_0^2<0$, an equilibrium state can
possibly exist only for $T<T_c$. For $T=T_c$ and $\omega_0\neq 0$, we
must have $I=0$.

\begin{figure}
\begin{center}
\includegraphics[clip,scale=0.3]{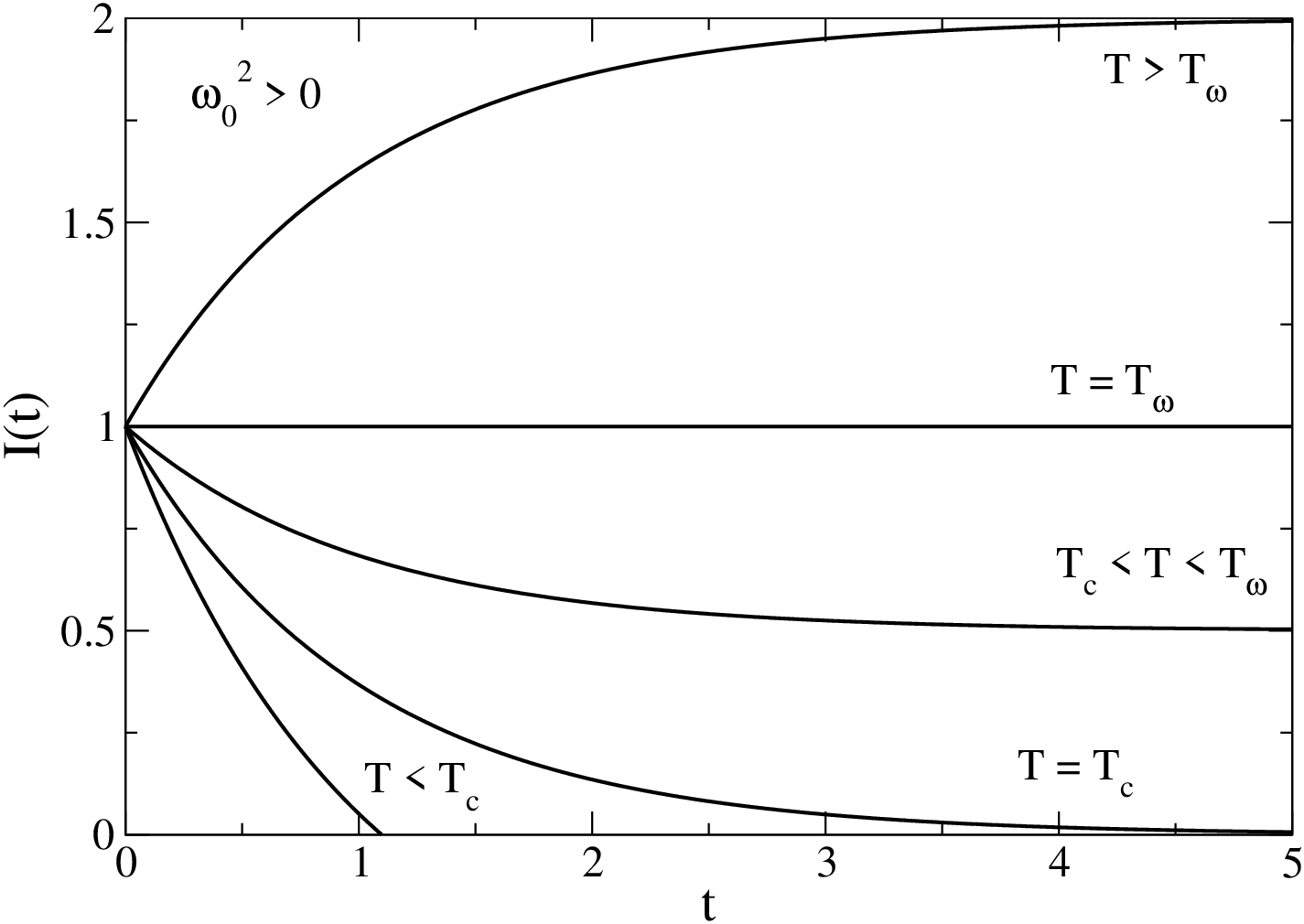}
\caption{Time evolution of the moment of inertia for an overdamped Brownian  system with logarithmic interactions and attractif harmonic potential.}
\label{hattractif}
\end{center}
\end{figure}

\begin{figure}
\begin{center}
\includegraphics[clip,scale=0.3]{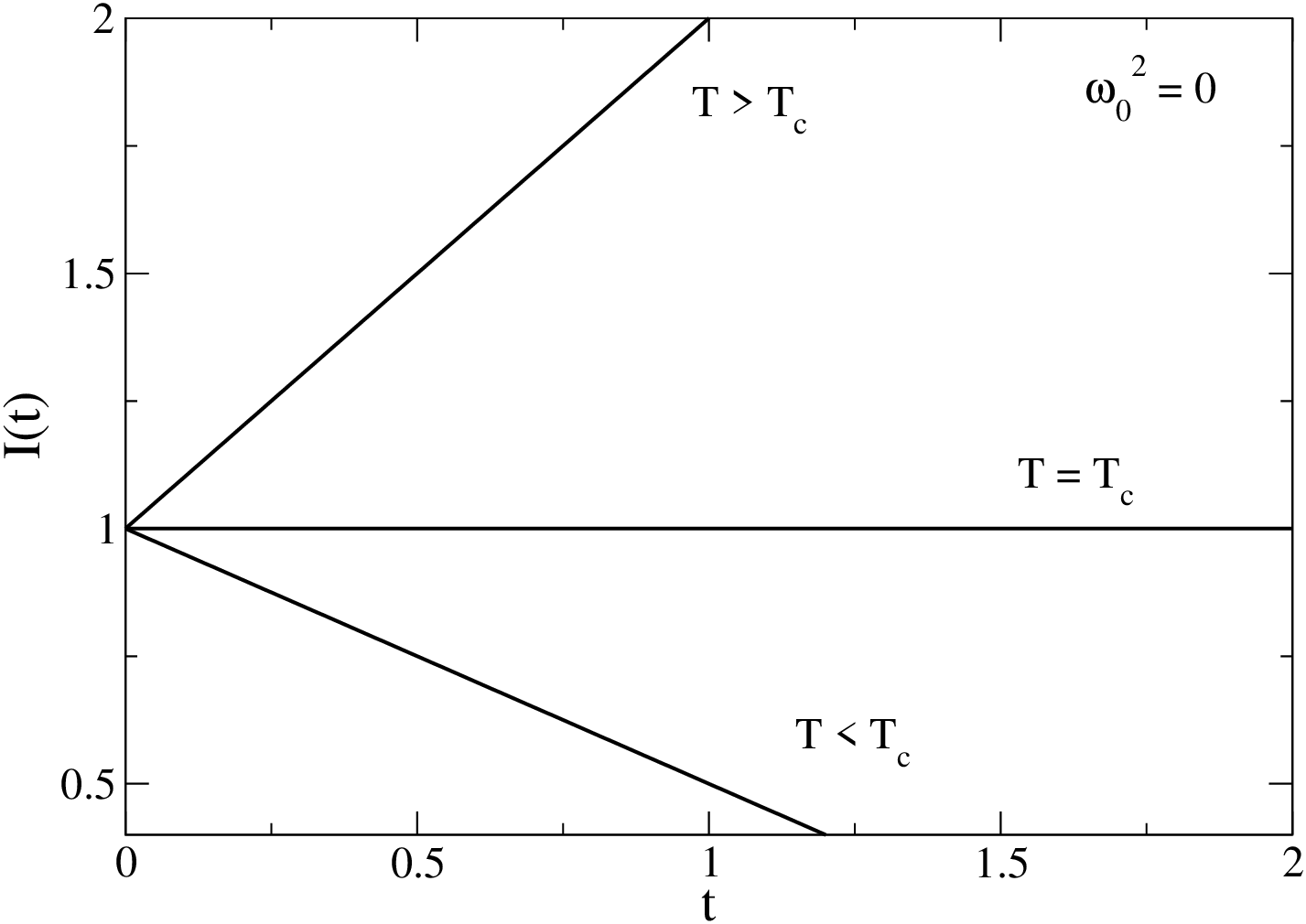}
\caption{Time evolution of the moment of inertia for an overdamped Brownian  system with logarithmic interactions.}
\label{hneutre}
\end{center}
\end{figure}

\begin{figure}
\begin{center}
\includegraphics[clip,scale=0.3]{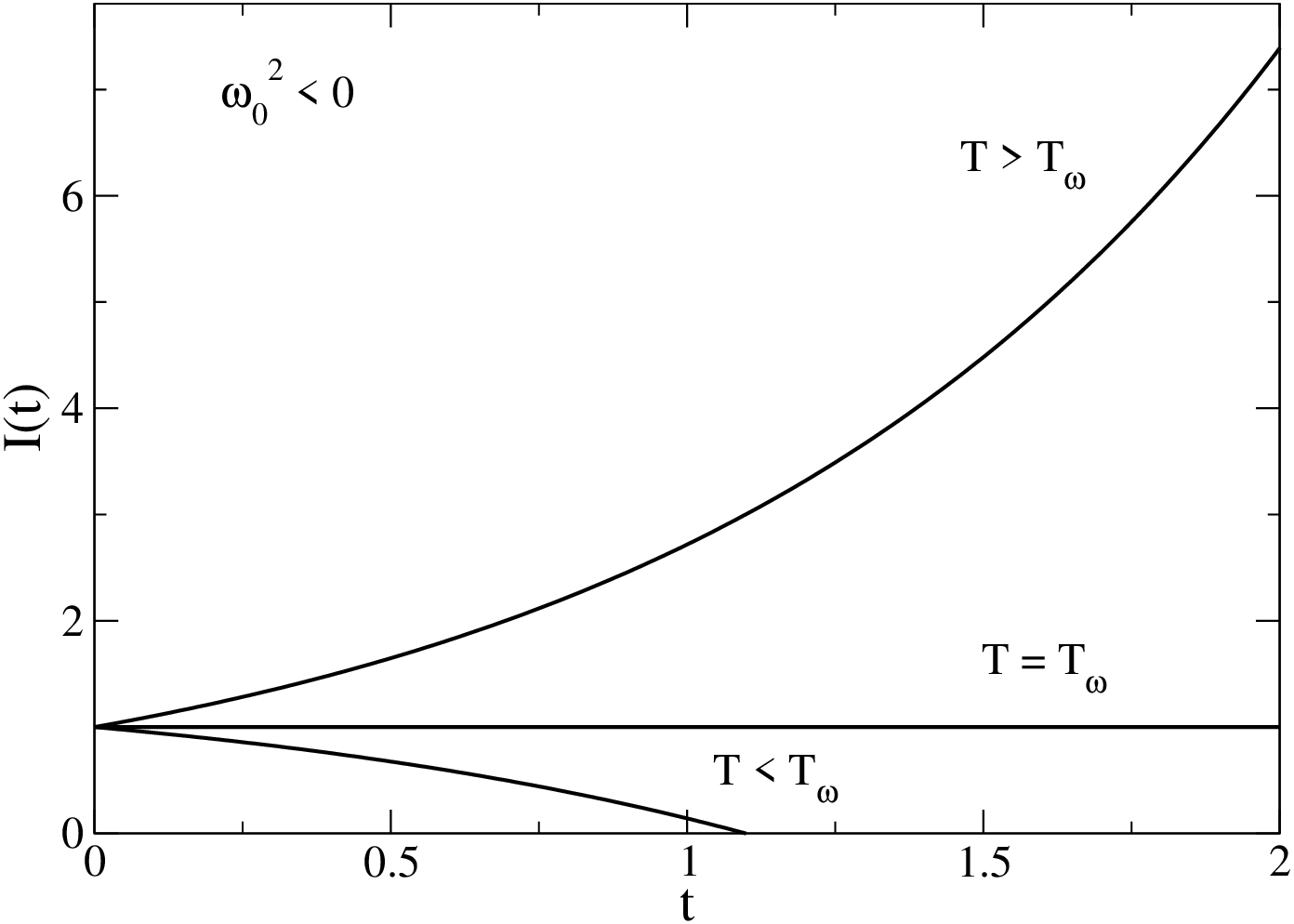}
\caption{Time evolution of the moment of inertia for an overdamped Brownian  system with logarithmic interactions and repulsive harmonic potential.}
\label{hrepulsif}
\end{center}
\end{figure}

We now consider the strong friction limit $\xi\rightarrow +\infty$
where inertial effects are negligible. In that limit, the velocities
thermalize on a timescale of order $1/\xi$. In that case, $K_{ij}
={1\over 2}Nk_{B}T\delta_{ij}$ and $P_{ij}=p\delta_{ij}$ with
$p=\sum_{s}\rho_{s}k_{B}T/m_{s}$ even if the system has not yet
reached a state of mechanical equilibrium \cite{virial}. From Eq. (\ref{virg5}), we obtain
the overdamped virial theorem for a self-gravitating Brownian gas
\begin{equation}
{1\over 2}\xi  {\dot I}_{ij}+\omega_0^2 I_{ij}=Nk_{B}T\delta_{ij}+W_{ij}-{1\over
2}\oint p(x_{i}dS_j+x_{j}dS_i).
\label{virg27}
\end{equation}
We can obtain this result in a different manner. In the strong
friction limit $\xi\rightarrow +\infty$, the inertial term in
Eq. (\ref{virg1}) can be neglected so that the stochastic equations of
motion reduce to
\begin{equation}
\dot x_{i}^{\alpha}=\mu_{\alpha}m_{\alpha}\sum_{\beta\neq \alpha}
{Gm_{\beta}(x_{i}^{\beta}-x_{i}^{\alpha})\over |{\bf
r}_{\beta}-{\bf
r}_{\alpha}|^{d+\gamma}}-\frac{\omega_0^2}{\xi}+\sqrt{2D'_{\alpha}}B_{i}^{\alpha}(t),
\label{virg28}
\end{equation}
where $D'_{\alpha}=k_{B}T\mu_{\alpha}$ is the diffusion coefficient is
physical space and $\mu_{\alpha}=1/(\xi m_{\alpha})$ the mobility. The
overdamped virial theorem (\ref{virg27}) can be directly obtained from
these stochastic equations \cite{virial}. The
scalar virial theorem reads
\begin{equation}
{1\over 2}\xi {\dot I} +\omega_0^2 I =dNk_{B}T+W_{ii} -dPV.
\label{virg29}
\end{equation}
For $\gamma=2-d$, using Eq. (\ref{virg15}), we
obtain
\begin{equation}
{1\over 2}\xi {\dot I} +{\omega_0^2 I}=dNk_{B}(T-T_{c})-dPV,
\label{virg30}
\end{equation}
with the exact critical temperature (\ref{virg23}). In an infinite domain ($P=0$), this relation reduces to
\begin{equation}
{1\over 2}\xi {\dot I} +{\omega_0^2 I}=dNk_{B}(T-T_{c}).
\label{virg31}
\end{equation}
This is a {\it closed} equation that can be solved analytically. For $\omega_0\neq 0$, the solution is
\begin{equation}
I(t)=\left\lbrack I(0)-\frac{dNk_B}{\omega_0^2}(T-T_c)\right\rbrack e^{-\frac{2\omega_0^2}{\xi}t}+\frac{dNk_B}{\omega_0^2}(T-T_c).
\label{virg31a}
\end{equation}
Let us introduce the new critical temperature 
\begin{equation}
k_B T_{\omega}=k_B T_c+\omega_0^2\frac{I(0)}{dN},
\label{rep1}
\end{equation}
depending on the initial value of moment of inertia (note that
$T_{\omega}>T_c$). This is the value at which the term in bracket in
Eq. (\ref{virg31a}) vanishes. If $T>T_c$, the system tends to an equilibrium state corresponding to $I_{eq}=\frac{dNk_B}{\omega_0^2}(T-T_c)$, see Eq. (\ref{virg26}). More precisely, for $T>T_{\omega}$, the moment of inertia increases, for $T_c<T<T_{\omega}$ it decreases and for $T=T_{\omega}$ it remains constant. If $T=T_c$, we find that $I(t)\rightarrow 0$ for $t\rightarrow +\infty$ implying a collapse in infinite time. If $T<T_c$, the moment of inertia vanishes at a time
\begin{equation}
t_{end}=\frac{\xi}{2\omega_0^2}\ln\left\lbrack 1+\frac{\omega_0^2 I(0)}{dNk_B (T_c-T)}\right\rbrack, 
\label{virg31b}
\end{equation}
implying the finite time collapse of the system (we recall that $I(t)=M\langle r^2\rangle(t)$). For $\omega_0=0$, we obtain
\begin{equation}
I(t)=\frac{2dNk_B}{\xi}(T-T_c)t+I(0).
\label{virg31c}
\end{equation}
If $T>T_c$, the system evaporates. If $T=T_c$, we find that $I(t)=I(0)$ for all times. If $T<T_c$, the moment of inertia vanishes at a time
\begin{equation}
t_{end}=\frac{\xi I(0)}{2dNk_B (T_c-T)},
\label{virg31d}
\end{equation}
implying the finite time collapse of the system. If $\omega_0^2=-\Omega_0^2<0$ (repulsive harmonic potential or rotation), the picture is different.   For $T>T_{\omega}$ (note that now
$T_{\omega}<T_c$), the system
evaporates. For $T<T_{\omega}$, which is possible iff $I(0)<dNk_B
T_c/(-\omega_0^2)$, the moment of inertia vanishes at a time
$t_{end}$, given by Eq. (\ref{virg31b}), implying finite time
collapse. Finally, for  $T=T_{\omega}$, the moment of inertia is
conserved. Some representative curves of these different evolutions are given in Figs. \ref{hattractif}, \ref{hneutre} and \ref{hrepulsif}.

We now give the proper form of virial theorem corresponding to the generalized Smoluchowski equation \cite{virial}:
\begin{equation}
\frac{\partial\rho}{\partial t}=\nabla\cdot\left\lbrack\frac{1}{\xi}\left (\nabla p+\rho\nabla\Phi+\rho\omega_0^2{\bf r}\right )\right\rbrack,
\label{virg32}
\end{equation}
where  the pressure $p=p({\bf r},t)$ is given by  an arbitrary  barotropic equation of state $p=p(\rho)$. We assume that the particles are subject to an external harmonic potential $V({\bf r})=\frac{1}{2}\omega_0^2 r^2$ and that they interact through an algebraic potential
\begin{equation}
\Phi({\bf r},t)=-\frac{G}{d+\gamma-2}\int \frac{\rho({\bf r}',t)}{|{\bf r}-{\bf r}'|^{d+\gamma-2}}\, d{\bf r}',
\label{virg33}
\end{equation}
if $\gamma\neq 2-d$, or a logarithmic potential
\begin{equation}
\Phi({\bf r},t)={G}\int {\rho({\bf r}',t)}\ln |{\bf r}-{\bf r}'|\, d{\bf r}',
\label{virg34}
\end{equation}
if $\gamma= 2-d$. The mean force of interaction acting on a particle
in ${\bf r}$ is
\begin{equation}
{\bf F}=-\nabla\Phi=-G\int \rho({\bf r}',t)\frac{{\bf r}-{\bf r}'}{|{\bf r}-{\bf r}'|^{d+\gamma}}\, d{\bf r}'.
\label{virg35}
\end{equation}
For simplicity, we assume that the particles have the same mass $m$. The Lyapunov functional associated with the Smoluchowski equation (\ref{virg32}) is the free energy
\begin{equation}
F=\int\rho\int_{0}^{\rho}\frac{p(\rho')}{\rho^{'2}}\, d\rho' d{\bf r}+\frac{1}{2}\int\rho\Phi\, d{\bf r}+\int\rho V\, d{\bf r},
\label{lyap1}
\end{equation}
and it satisfies an $H$-theorem, i.e. $\dot F\le 0$. The Smoluchowski
equation (\ref{virg32}) with an isothermal equation of state $p({\bf
r},t)=\rho({\bf r},t) k_B T/m$ is the mean field Fokker-Planck
equation associated with the overdamped stochastic process
(\ref{virg28}). The potential energy tensor is defined by
\begin{equation}
W_{ij}=-\int\rho x_i \frac{\partial\Phi}{\partial x_j}\, d{\bf r},
\label{virg36}
\end{equation}
while the virial is
\begin{equation}
W_{ii}=-\int\rho {\bf r}\cdot\nabla\Phi\, d{\bf r}.
\label{virg36b}
\end{equation}
Substituting Eq. (\ref{virg35}) in Eq. (\ref{virg36}) and using the usual symmetrization procedure, it can be rewritten
\begin{equation}
W_{ij}=-\frac{1}{2}G\int\rho({\bf r})\rho({\bf r}') \frac{(x_i - x_i')(x_j - x_j')}{|{\bf r}-{\bf r}'|^{d+\gamma}}\, d{\bf r}d{\bf r}'.
\label{virg37}
\end{equation}
Contracting the indices, we get
\begin{equation}
W_{ii}=-\frac{1}{2}G\int \frac{\rho({\bf r})\rho({\bf r}')}{|{\bf r}-{\bf r}'|^{d+\gamma-2}}\, d{\bf r}d{\bf r}'.
\label{virg38}
\end{equation}
For $\gamma\neq 2-d$, we obtain
\begin{eqnarray}
W_{ii}=(d+\gamma-2)W,
\label{virg39}
\end{eqnarray}
where $W=\frac{1}{2}\int\rho\Phi\, d{\bf r}$ is the mean field potential energy. For   $\gamma= 2-d$, we obtain
\begin{eqnarray}
W_{ii}=-\frac{GN^2m^2}{2}.
\label{virg40}
\end{eqnarray}
Introducing the moment of inertia tensor
\begin{equation}
I_{ij}=\int\rho x_ix_j\, d{\bf r},
\label{virg41}
\end{equation}
we find that the tensor virial theorem associated with the generalized Smoluchowski equation (\ref{virg32}) is given by
\begin{equation}
{1\over 2}\xi  {\dot I}_{ij}+\omega_0^2 I_{ij}=\delta_{ij}\int p\, d{\bf r}+W_{ij}-{1\over
2}\oint p(x_{i}dS_j+x_{j}dS_i).
\label{virg42}
\end{equation}
The scalar virial theorem, obtained by contracting the indices, takes the form
\begin{equation}
{1\over 2}\xi {\dot I}+\omega_0^2 I =d\int p\, d{\bf r}+W_{ii} -dPV,
\label{virg43}
\end{equation}
where 
\begin{equation}
I=\int \rho r^2\, d{\bf r},
\label{virg44}
\end{equation}
is the moment of inertia. The equilibrium scalar virial theorem is 
\begin{equation}
\omega_0^2 I =d\int p\, d{\bf r}+W_{ii} -dPV.
\label{virg45}
\end{equation}
For an isothermal equation of state $p=\rho k_B T/m$, we recover Eq. (\ref{virg29}) where $W_{ii}$ is now given by Eq. (\ref{virg40}).
For a logarithmic potential ($\gamma=2-d$), we recover Eq. (\ref{virg30}) where $T_c$ is given by 
\begin{equation}
k_{B}T_{c}={GNm^{2}\over 2d}.
\label{virg46}
\end{equation}
At equilibrium, we recover Eq. (\ref{virg21}).

Finally, we give the proper form of virial theorem for the damped barotropic Euler equations 
\begin{equation}
\frac{\partial\rho}{\partial t}+\nabla\cdot (\rho {\bf u})=0,
\label{virg47}
\end{equation}
\begin{equation}
\frac{\partial {\bf u}}{\partial t}+({\bf u}\cdot\nabla){\bf u}=-\frac{1}{\rho}\nabla p-\nabla\Phi-\xi {\bf u}-\omega_0^2{\bf r},
\label{virg48}
\end{equation}
under the same conditions as before. The tensor virial theorem is given by
\begin{eqnarray}
{1\over 2} {\ddot I}_{ij}+{1\over 2}\xi  {\dot
I}_{ij}+\omega_0^2 I_{ij} =\int\rho u_i u_j\, d{\bf r}+\delta_{ij}\int p\, d{\bf r}\nonumber\\
+W_{ij} -{1\over 2}\oint
p(x_{i}dS_j+x_{j}dS_i),
\label{virg49}
\end{eqnarray}
and the scalar virial theorem by
\begin{eqnarray}
{1\over 2} {\ddot I}+{1\over 2}\xi  {\dot
I}+\omega_0^2 I =\int\rho {\bf u}^2\, d{\bf r}+d\int p\, d{\bf r}+W_{ii} -dPV. \nonumber\\
\label{virg50}
\end{eqnarray}
At equilibrium, we obtain Eq. (\ref{virg45}). The virial theorem for the barotropic Euler equations is recovered by taking $\xi=0$ \cite{chandravirial,bt}.
 The Lyapunov functional associated with the damped Euler equations (\ref{virg47})-(\ref{virg48}) is the free energy
\begin{equation}
F=\int\rho\int_{0}^{\rho}\frac{p(\rho')}{\rho^{'2}}\, d\rho' d{\bf r}+\frac{1}{2}\int\rho\Phi\, d{\bf r}+\int\rho V\, d{\bf r}+\int \rho \frac{{\bf u}^2}{2}\, d{\bf r},
\label{lyap2}
\end{equation}
and it satisfies an $H$-theorem, i.e. $\dot F\le 0$ if $\xi\neq 0$. For the Euler equations ($\xi=0$), the energy functional (\ref{lyap2}) is conserved  $\dot F=0$.

\end{document}